\documentclass[12pt,english]{article}
\usepackage[english]{babel}
\usepackage{amsmath,amssymb,amscd}
\usepackage{amsfonts}
\usepackage{graphicx}
\usepackage{appendix}

\makeatletter
\renewcommand\section{\@startsection {section}{1}{\z@}%
                                   {-5.5ex \@plus -1ex \@minus -.2ex}%nn
                                   {2.3ex \@plus.2ex}%
                                   {\normalfont\large\bfseries}}
\renewcommand\subsection{\@startsection{subsection}{2}{\z@}%
                                     {-3.25ex\@plus -1ex \@minus -.2ex}%
                                     {1.5ex \@plus .2ex}%
                                     {\normalfont\bfseries}}

\numberwithin{equation}{section}

\makeatother

\date{}

\newcommand{\ba}{\begin{align}}

\newcommand{\bea}{\begin{eqnarray}}
\newcommand{\eea}{\end{eqnarray}}
\newcommand{\be}{\begin{equation}}
\newcommand{\ee}{\end{equation}}
\newcommand{\eq}[1]{(\ref{#1})}

\newcommand{\Z}{{\mathbb Z}}

\newcommand{\cO}{{\mathcal O }}

\newcommand{\cD}{{\mathcal D }}

\newcommand{\ie}{i.e.~}

\newcommand{\ma}{{\textrm{ch}}}
\newcommand{\mab}{{\mathbf{ch}}}

\newcommand{\ml}{\chi}
\newcommand{\mlb}{\boldsymbol{\chi}}

\def\transpose{\top}
\def\Nb{\mathbf{N}}
\def\Gb{\mathbf{G}}
\def\fb{\mathbf{f}}
\def\Mb{\mathbf{M}}
\def\Sb{\mathbf{S}}
\def\Ab{\mathbf{A}}
\def\Fb{\mathbf{F}} % DF 2013-07-18

\def\Db{\mathbf{D}}
\def\Yb{\mathbf{Y}}
\def\tot{\mathrm{tot}}

\def\eq{\begin{equation}}
\def\en{\end{equation}}
\def\eqa{\begin{eqnarray}}
\def\ena{\end{eqnarray}}
\def\aeq#1{\begin{align}#1\end{align}}  % for amsmath
\def\ignorethis#1{}
\def\Integers{\mathbb{Z}}		%Integers
\def\tr{\mathop{\mathrm{tr}}}
\def\max{\mathit{max}}
\def\Kb{\mathbf{K}}
\def\odd{\mathit{odd}}
\def\Deltamax{\Delta_{B}}
\def\nD{n_{{\cD }}}
\def\bo{\Deltamax}
\def\htot{h^\tot}
\def\Ysol{\Yb_{\!\!\mathit{sol}}}
\def\Dsol{\cD_{\!\!\mathit{sol}}}

\title{\vspace{-1cm}\begin{flushright}{\small RUNHETC-2013-16}\end{flushright}\vspace{2cm}
\LARGE Constraints on 2d CFT partition functions
}

\author
{
Daniel Friedan$^{1,2}$\footnote{friedan@physics.rutgers.edu}\,,\ Christoph A.~Keller$^1$\footnote{keller@physics.rutgers.edu}
\\
\\
$^1$NHETC and Department of Physics and Astronomy\\
Rutgers, The State University of New Jersey\\
Piscataway, New Jersey 08854-8019, USA.\\
\\
$^2$The Science Institute, The University of Iceland,
Reykjavik, Iceland
}

\begin{document}

\maketitle

\begin{center}
{\bf Abstract}
\end{center}

Modular invariance is known to constrain the spectrum of 2d conformal
field theories.  We investigate this constraint systematically, using
the linear functional method to put new improved upper bounds on the
lowest gap in the spectrum.  We also consider generalized partition
functions of $N = (2,2)$ superconformal theories and discuss the
application of our results to Calabi-Yau compactifications.  
For Calabi-Yau threefolds with no enhanced symmetry
we find that there must always be non-BPS primary states of
weight $0.6$ or less.
\newpage

\section{Introduction}
The conformal bootstrap is the project of
constructing conformal field theories from 
consistency conditions imposed by conformal invariance 
\cite{Polyakov:1970xd,Migdal:1972tk,Polyakov:1974gs}.
Historically it has proved very powerful
in the analysis of two dimensional conformal field theories with 
central charge $c<1$ \cite{Belavin:1984vu,DiFrancesco:1997nk}.
Compared to other methods, its main advantage is that it does
not rely on a Lagrangian prescription of the CFT.
It makes use of the conformal symmetry
of the theory by decomposing amplitudes into conformal blocks ---
the contributions of the irreducible representations of the conformal 
group.
In principle it is thus possible to classify and construct
all CFTs, including strongly coupled ones.

More recently the work of \cite{Rattazzi:2008pe} sparked renewed interest
in the bootstrap approach. In it the authors derive constraints
on the spectrum of CFTs from the condition of crossing
symmetry of the four point function. They use 
explicit expressions for the conformal blocks
derived in \cite{Dolan:2000ut,Dolan:2003hv}.
From these they numerically derive a vector
space of constraints, and optimize over the constraints
to obtain upper bounds on the dimensions of fields.

Although these methods apply to CFTs in any dimension,
the situation in two dimensions is special.
The finite dimensional group of global conformal symmetries --- or, rather, its Lie algebra 
--- is enhanced to the infinite dimensional Lie algebra of holomorphic and 
anti-holomorphic local conformal transformations.
The quantum operators representing these maps
form two commuting Virasoro algebras.
Their quantum central charge $c$
measures, roughly speaking, the number of degrees 
of freedom in the theory.
For $c<1$, bootstrap methods
are very powerful for the classification of conformal field theories.  
For $c\geq1$, unfortunately, the expressions for the conformal blocks of the
Virasoro algebra are much more complicated, 
so that the bootstrap method using the full Virasoro algebra has not been
practical. For this reason the general $c\geq 1$ picture 
is still unknown, though many 
explicit examples are known.

In two dimensions we require that the CFT be
consistent not just on the sphere, but on arbitrary Riemann 
surfaces.
This \emph{modular invariance} condition
was first discovered in the context of string theory 
\cite{Shapiro:1972ph}.
Of course analogous conditions arise in higher dimension, but little 
or nothing is known about their significance.  In two dimensions, 
modular invariance is known 
to be essential \cite{Cardy:1986ie}.
It turns out that the necessary and sufficient conditions for the 
theory to be defined consistently on all two dimensional surfaces
are: (1) crossing symmetry of the four-point functions on the sphere,
and (2) modular invariance of the partition function and 
the one-point functions on the torus \cite{Moore:1988qv}.
Higher genus amplitudes can then be constructed by gluing various punctured spheres
and tori together.  The above conditions
ensure that this procedure gives consistent answers.

In this paper we will focus solely on 
modular invariance of the torus correlation function
with no operators inserted.
The idea is to apply methods used in the modern 
bootstrap to this amplitude, as
has been done in \cite{Hellerman:2009bu,Hellerman:2010qd,Friedan:2012jk,Keller:2012mr,Friedan:2013bha}.
The big advantage compared to the study of correlation functions (as 
in \cite{Rattazzi:2008pe}) is that the expressions
for the contributions of the various representations
are known exactly and have a relatively simple form.
Eventually one will want to combine these results 
with results from one-point functions on the torus and from 
four-point functions on the sphere,
something we will leave for future work.

The amplitude of a CFT on a torus with no insertion of
operators can be written as the
partition function of the quantum field theory living on the space $S^1$,
\eq
Z(\tau,\bar\tau) = \tr \left ( e^{2\pi i \tau (L_{0}-c/24)} e^{-2\pi i \bar 
\tau (\bar L_{0}-c/24)}\right )\,,
\label{eq:partfn1}
\en
where $\tau$, the modulus of the torus, is in the complex upper half plane.
The torus is the complex plane modulo the lattice $\{m+n\tau: 
m,n\in\Integers\}$.
The operators $L_{0}$, $\bar L_{0}$ are the middle elements of the 
two Virasoro algebras.
The hamiltonian is $H=2\pi(L_{0}+\bar L_{0})$ while 
$P=2\pi(L_{0}-\bar L_{0})$
generates translations in space, the circle $S^{1}$.
Since the amplitude can only depend on the complex structure of the torus,
it follows that it has to be invariant under the action
of the modular group $SL(2,\Z)$, 
since such transformations give conformally equivalent tori.
The modular group is generated by the  
transformations  $T: \tau\rightarrow \tau+1$ and $S: \tau\rightarrow 
-1/\tau$, 
and modular invariance is hence equivalent to invariance under the 
transformations $S$ and $T$.

Irreducible representations of the Virasoro symmetry are labelled
by their right and left conformal weights $(h,\bar h)$.
The quantization on $S^{1}$ can be interpreted as the radial quantization of the 
euclidean CFT.
The irreducible representations correspond to primary fields of 
dimension $h+\bar h$ and spin $h-\bar h$.
The partition function can be expanded in characters of the 
irreducible representations,
\be
Z(\tau,\bar\tau)
=  \sum_{(h,\bar h)} 
\overline{\chi_{\bar h}(\tau)}\, N_{\bar h h}\,\chi_{h} (\tau) \, .
\label{eq:partfnexpanded1}
\ee
The multiplicity $N_{\bar h h}$ counts the number of times
that the representation $(h,\bar h)$ occurs in the spectrum,
and hence is a nonnegative integer.
The function $\chi_{h} (\tau)$ is the character of the 
representation of a single 
Virasoro algebra,
\be
\chi_{h} (\tau) = \tr \left ( e^{2\pi i \tau (L_{0}-c/24)}\right )\,,
\ee
where the trace is over the irreducible representation of weight $h$.
(In $N_{\bar h h}$, we write the subscripts in the order $\bar h h$ because we
regard $N_{\bar h h}$ as a hermitian form on the space of characters.)

At the core of the bootstrap approach is the observation
that the characters
$\chi_h(\tau)$ themselves are not modular invariant. This means
that only very specific choices for the multiplicities
$N_{\bar h h}$ lead to an invariant partition function.
Invariance under $T: \tau\rightarrow \tau+1$ is equivalent
to imposing integer spin $h-\bar h \in \Integers$. The more interesting
constraint is invariance under
$S: \tau\rightarrow -1/\tau$,
\be\label{intromain}
Z(\tau,\bar\tau) = Z(-1/\tau,-1/\bar\tau)\,.
\ee
Since (\ref{intromain}) must hold for all values
of $\tau$, a priori it gives an infinite number
of constraints on the multiplicities $N_{\bar h h}$.

One way to express these constraints is via linear functionals $\rho$
on the space of functions of $\tau$ \cite{ElShowk:2012hu}.
Every linear functional $\rho$ acts on (\ref{intromain})
to give, by (\ref{eq:partfnexpanded1}), a linear constraint on the 
$N_{\bar h h}$.
If we choose a suitable infinite set of linear functionals,
\ie a basis of the dual space of the functions of $\tau$,
we can in principle recover the full set of constraints.
To obtain a complete classification of all allowed spectra
$N_{\bar h h}$, one would have to determine all solutions
to this system of constraints. With our current understanding
this is not practical, and we will not attempt to do so.
Instead, following \cite{Hellerman:2009bu}, we will pursue the more modest goal of putting
bounds on basic features of the spectrum.
Specifically, we investigate the lowest gap $\Delta_1$ in the spectrum
---
the value of the total conformal 
weight $\Delta_1=h+\bar h $ of the lowest lying non-vacuum primary field.
It turns out that $\Delta_1$ cannot be too big, as
otherwise it becomes impossible to satisfy (\ref{intromain}).
The goal of our work is to determine an upper
bound $\bo$ on $\Delta_1$. 
The approach is to consider some
$n$-dimensional
subspace of linear functionals, and find the linear functional
in this space that gives the strongest bound.
In principle, the optimal bound is obtained in the limit 
$n\rightarrow \infty$.
But as the complexity of the computation grows with $n$,
in practice we are limited by our computing power
to relatively modest values of $n$.

In the first half of this paper we consider bosonic conformal field
theories, in which the conformal symmetry is the Virasoro algebra.
We calculate bounds $\bo(c)$
that depend on the central charge $c$.
In the second half of this paper we consider
models with extended $N=(2,2)$ supersymmetry.
Specifically we are interested in non-linear sigma models whose
target space is a Calabi-Yau manifold. In this case the central 
charge
is fixed by the complex dimension $d$ of the Calabi-Yau manifold,
$c=3d$. 
Because of the $N=(2,2)$ supersymmetry, the spectrum of this
theory contains BPS states. In general a lot is known about the BPS spectrum,
which is related to the topology
of the Calabi-Yau, namely its Hodge numbers and elliptic genus.
The linear functional method can give information on the non-BPS states,
about which much less is known. In principle sufficient knowledge
of the non-BPS states should allow to answer questions on the
geometry of the manifold. If for instance one could show that
a given set of topological numbers cannot lead to a consistent non-BPS spectrum,
then this would rule out the existence of a Calabi-Yau manifold with 
such topology.
In particular this could help answer the still open question,
whether there are only finitely many topological families of
Calabi-Yau threefolds \cite{yau,Reid}.
As it turns out, the methods of the present paper cannot produce 
strong enough constraints on the non-BPS spectrum to yield answers to 
such geometric questions.

In \cite{Keller:2012mr}, the linear functional method was used to find a bound $\bo$ on the lowest lying non-BPS state
for Calabi-Yau threefolds.
The bound is
a function of the Hodge numbers of the Calabi-Yau.
The bound was produced using a two dimensional subspace of linear 
functions.
To improve on this calculation, we find it useful to express the
generalized $N=2$ characters in theta functions.
The calculation turns out to simplify dramatically for Calabi-Yau threefolds.
Making the technical assumption that the theory does not 
have an enhanced symmetry beyond the extended $N=(2,2)$,
we improve significantly on the bound of~\cite{Keller:2012mr} --- see
figure~\ref{Fig:N2diffops}. In particular we
find that there is always a non-BPS state of total
weight $\Delta$ less than $\bo=0.6$.  (We include among
the non-BPS states any pairs of BPS states that can
combine to form non-BPS states.)
We present evidence that our results are close to optimal
for the linear functional method.

We adopt a number of limitations on our project to make the computations more tractable.
First, we restrict our attention to theories without enhanced 
symmetry, \ie without additional holomorphic or anti-holomorphic fields.
In the bosonic case, we assume there is only the Virasoro algebra.
Also, we assume $c>1$, leaving out the case 
$c=1$.
In the $N=(2,2)$ case, we assume there is only the extended N=2 algebra.
These restrictions allow us to avoid complications due to degenerate 
representations of the symmetry algebra.
Second, following \cite{Hellerman:2009bu}, 
we restrict the partition function to purely imaginary values 
of $\tau$.
The partition function then depends only on the spectrum of total 
conformal weights $\Delta= h+\bar h$.
Information about the spins $h-\bar h$ is not used.
In principle it is straightforward to apply the linear functional 
method for complex $\tau$, but our implementation of the method
cannot handle the condition of $T$ invariance --- that the spins $h-\bar h$
must be integers (or half-integers for fermions).
Our bounds would be strengthened if we could enforce $T$ invariance.
Third, although the multiplicities $N_{\bar h  h}$ are integers,
the linear functinal method treats them as nonnegative \emph{real} numbers.
If we could use the constraint that the multiplicities are integers, 
we could strengthen the bounds.
Unfortunately, it is very hard to impose integrality with our methods.

Finally, we are only checking consistency of the 0-point
function on the torus. We should be
able to treat the torus 1-point with similar methods, while the 4-point function on the
sphere can be dealt with using the ordinary bootstrap
methods. We then expect the challenging part to be to
combine those various types of results to obtain 
overall constraints on the theory. Such an approach could yield significant 
improvements over our current bounds.

\section{General setup}

\subsection{The partition function}
Let us explain how the linear functional method can
be used to find constraints on the spectrum.
We base our exposition of the abstract linear functional method on \cite{ElShowk:2012hu}.
The method can give negative answers
to questions of the form:
\begin{quote}
For a given value of the central charge $c$,
does there exist a modular invariant partition function
whose conformal weights $(h,\bar h)$ lie in a given set $S$?
\end{quote}
If there is no modular invariant partition function,
then there can be no CFT whose conformal weights lie in $S$.
In practice, the set $S$ will be of the form
\be
S_{\Delta_{1}} = \left\{ (h,\bar h) \, : \,
h,\bar h \ge 0 , \;
h + \bar h \ge \Delta_{1}\right\}
\label{eq:SDelta1}
\ee
for some $\Delta_{1}>0$.  
To say that all the conformal weights lie in $S_{\Delta_{1}}$ is to say 
that all the scaling dimensions $\Delta=h+\bar h$ are $\ge 
\Delta_{1}$.
The lowest gap in the spectrum is $\Delta_{1}$.
A negative answer to the question means that
$\bo=\Delta_{1}$ is an upper bound on the gap.  Every CFT with central charge $c$ must 
have at least one scaling dimension  $\Delta\le\bo$. 
Our goal is the strongest
such bound, the lowest value we can find for $\bo$.

We use the Virasoro symmetry.
The unique ground state of the CFT
generates a unique
representation with $(h,\bar h)=(0,0)$, which by convention we omit 
from the set $S$.
The expansion~(\ref{eq:partfnexpanded1})  of the partition 
function in Virasoro characters
separates into a known contribution $Z_{0,0}$ from the ground state
representation plus the remaining sum,
\be
Z = Z_{0,0} + \sum_{(h,\bar h)\in S} 
N_{\bar h h} Z_{\bar h,h}\
\,,\qquad
Z_{\bar h,h} = \overline{\chi_{\bar h}( \tau)} \chi_{h} (\tau)
\,.
\label{eq:partfnexpanded2}
\ee
The sum contains the unknowns 
of the problem, namely the numbers $N_{\bar h h}$ which specify the conformal weights 
that occur in the spectrum, and their multiplicities.

Let $P_{\odd}$ be the projection on functions 
which are odd under $\tau\rightarrow -\tau^{-1}$,
\be
P_{\odd}f(\tau,\bar \tau) =: \frac12 \left [ f(\tau,\bar \tau) - 
f(-\tau^{-1},-\bar \tau^{-1})\right ]
\, .
\ee
Modular invariance under $\tau\rightarrow 
-\tau^{-1}$ can then be rewritten
\be
P_{\odd} Z =0\,.
\ee
In more general terms, $P_{\odd}$ projects on a subspace of functions 
complementary to the invariant functions.
Combining with~(\ref{eq:partfnexpanded2})
gives the central equation for our analysis,
\be
- P_{\odd} Z_{0,0} = \sum_{(h,\bar h)\in S} N_{\bar h h} P_{\odd}Z_{\bar h,h}
\,.
\label{eq:modinv}
\ee
Existence of a modular invariant partition function is equivalent to 
existence of a set of multiplicities 
$N_{\bar h h}$ satisfying~(\ref{eq:modinv}).

To simplify the problem we now abandon the integrality constraint on the multiplicities, allowing
the $N_{\bar h h}$ to be arbitrary nonnegative \emph{real} 
numbers.
If there is no solution to~(\ref{eq:modinv}) with nonnegative
real $N_{\bar h h}$, then there is certainly no solution with
nonnegative integer $N_{\bar h h}$.  Of course the converse is not
true. 

The functions that can appear on the rhs of~(\ref{eq:modinv})
now form a convex cone
\be
C_{S} = \bigg\{\sum_{(h,\bar h)\in S} N_{\bar h h} P_{\odd}Z_{\bar 
h,h} \, : N_{\bar h h} \geq 0 \bigg\}
\ee
within the vector space of real analytic functions of $\tau$ that are 
odd under $\tau\rightarrow -\tau^{-1}$.
The left hand side of~(\ref{eq:modinv}) is a vector in this function space,
\be
v_{0} = - P_{\odd} Z_{0,0}\,.
\ee
There exists at least one solution of~(\ref{eq:modinv})
with real multiplicities iff $v_{0}\in C_{S}$.
There is no real solution of~(\ref{eq:modinv}) iff 
$v_{0}\notin C_{S}$. The original
problem is reduced to checking if the vector $v_0$ is in the cone
$C_S$.

The sets $S_{\Delta_{1}}$ defined in~(\ref{eq:SDelta1}) become 
smaller as $\Delta_{1}$ increases,
so the cones $ C_{\Delta_{1}} =C_{S_{\Delta_{1}}}$
become narrower.
It will turn out that,
for sufficiently 
small $\Delta_{1}$,
$v_{0}$ always lies in the interior of $C_{\Delta_{1}}$.
So there is always a real solution of~(\ref{eq:modinv}) for sufficiently small $\Delta_{1}$.
As $\Delta_{1}$ increases, the cone $ C_{\Delta_{1}}$ narrows.
At a certain value of $\Delta_{1}$, the boundary of the cone will hit the 
vector $v_{0}$.
For larger values of $\Delta_{1}$, the vector $v_{0}$ lies outside the 
cone.
The best upper bound $\bo$ is the value of $\Delta_{1}$ 
where the boundary of the cone $ C_{\Delta_{1}}$
hits $v_{0}$.

\subsection{The linear functional method}
The linear functional method is based on the fact that
$v_0 \notin C_S$ if
there is a hyperplane in the function space
that separates the vector $v_{0}$ from the cone $C_{S}$.
In fact, the converse is also true.
If $v_0 \notin C_S$ then there exists a 
separating hyperplane, as follows from 
the generalized Farkas Lemma~\cite{Farkas}.
\begin{figure}[htbp]
\begin{center}
\includegraphics[height=.3\textwidth]{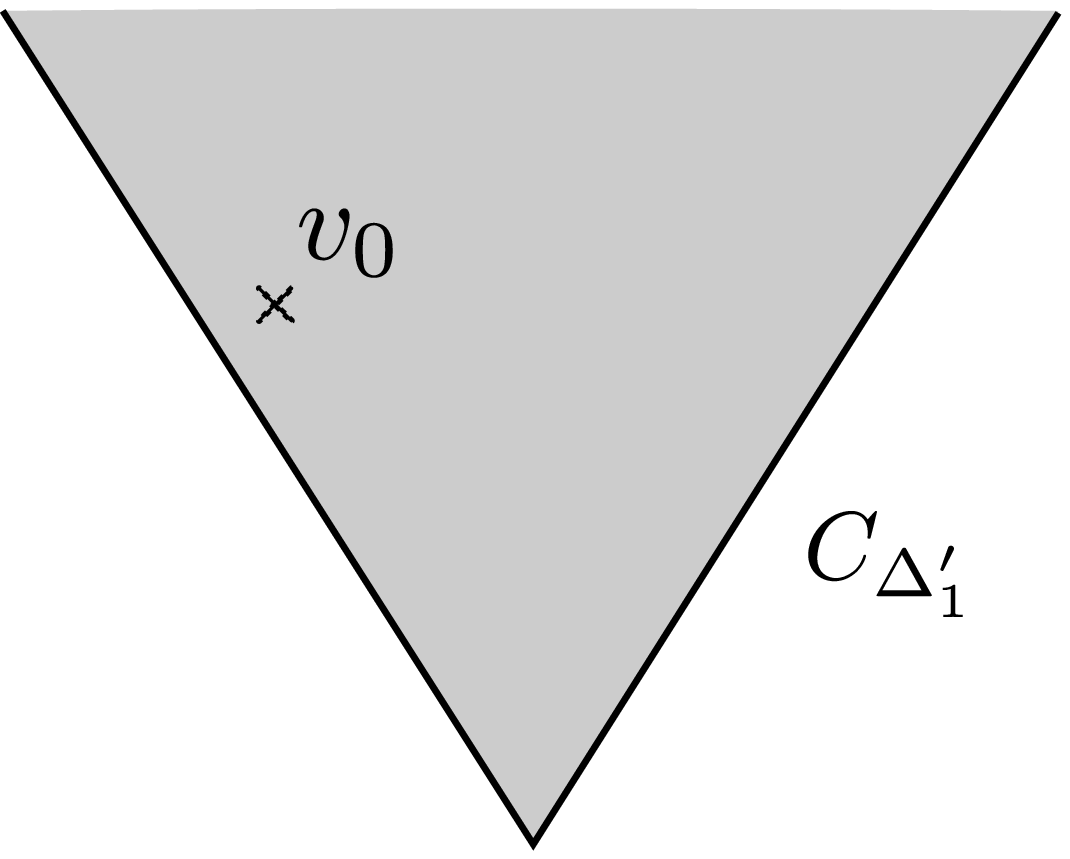}
\includegraphics[height=.3\textwidth]{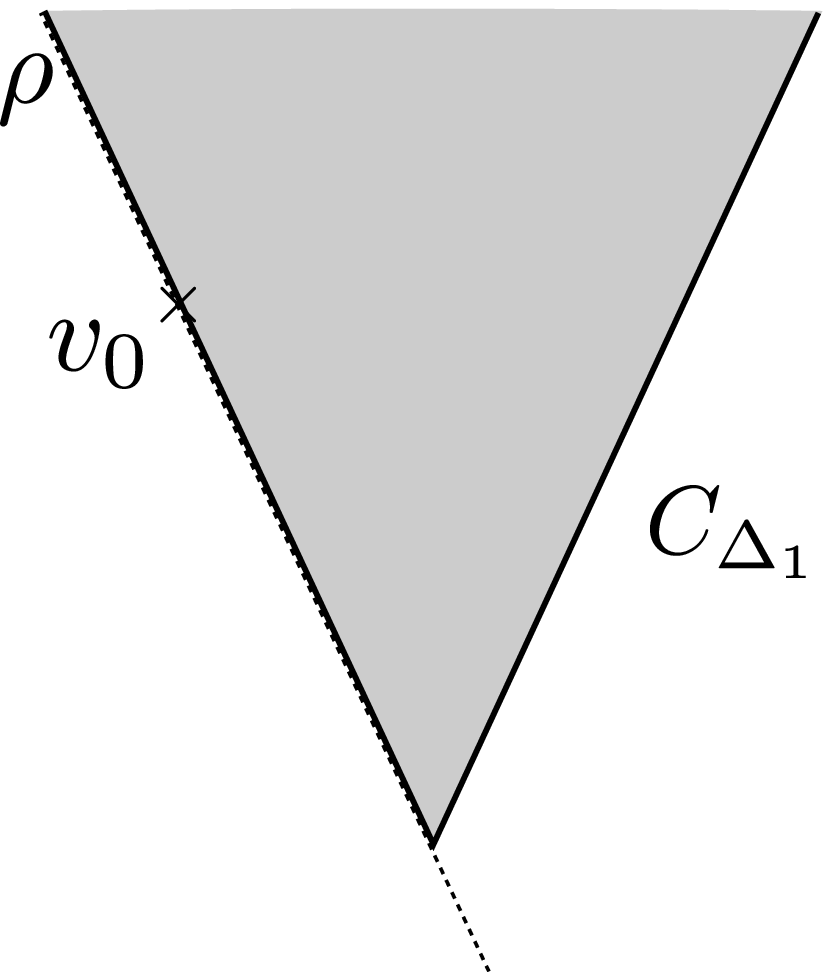}
\includegraphics[height=.3\textwidth]{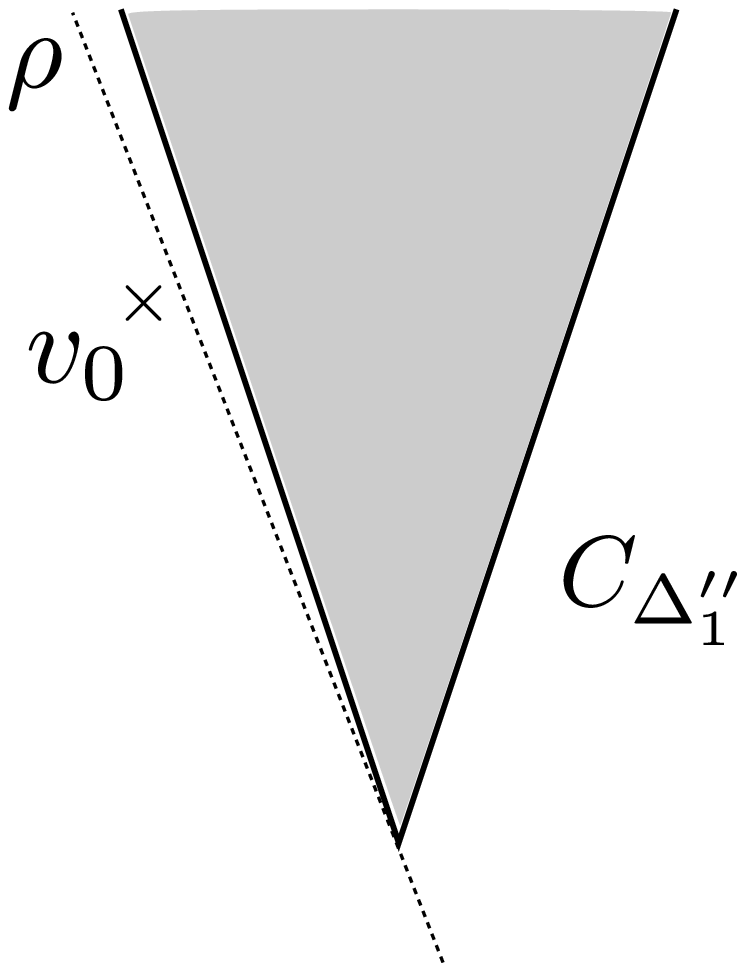}
\caption{The situation for $\Delta_1' < \Delta_1 < \Delta''_1$: For 
$\Delta_1'$, $v_{0}$
is within the cone $C_{\Delta'_1}$ so it is impossible to find a
separating plane. For $\Delta_1''$, $v_{0}$ is outside of the cone, and we
can find a separating plane $\rho$.
}
\label{Fig:cone3}
\end{center}
\end{figure}
Hyperplanes are in 1-to-1 correspondence with nonzero linear 
functionals $\rho$ modulo scaling $\rho\rightarrow \kappa \rho$,
$\kappa\ne 0$.
The hyperplane is the kernel of $\rho$.
It separates $v_{0}$ from $C_{S}$ iff 
(after scaling $\rho$ by $\pm 1$)
\be
\rho(C_{S})\subset [0,\infty)
\,\text{ and }\,
\rho(v_{0}) <0\,,
\ee
\ie $C_{S}$ is on one side 
of the hyperplane, where $\rho\ge 0$,
and $v_{0}$ is on the other side, where $\rho<0$.

Note that we can settle the question of existence of a 
separating hyperplane by
solving an optimization problem.
We maximize the objective function
\be
\cO = -\rho(v_{0})=\rho(P_{\odd} Z_{0,0})
\label{eq:optprob1}
\ee
over all linear functionals $\rho$ satisfying the semidefinite condition
\be
\rho(C_{S}) \subset [0,\infty)\,.
\label{eq:optprob2}
\ee
When the result is $\cO_{\max}>0$, then any solution $\rho$ of the optimization 
problem gives a separating hyperplane.  When $\cO_{\max}\le 0$, there is no separating hyperplane.
When the result is $\cO_{\max} = 0$, the vector $v_{0}$ lies just
on the boundary of the cone $C_{S}$,
giving our bound $\Delta_{B}$.

In the optimization problem,
the condition that $\rho$ be positive semidefinite on the cone 
$C_{S}$ is equivalent to the collection of inequalities
\be
\rho(P_{\odd}Z_{\bar h,h}) \ge 0\,, \quad  (h,\bar h) \in S\,.
\ee
The practical difficulty in using the linear functional method is to 
find an effective means of
enforcing the semidefinite condition.

\subsection{Reduced partition function}
Before discussing explicit linear functionals, let us first
simplify the expressions a bit by using the representation
theory of the Virasoro algebra.
The Virasoro characters for bosonic CFTs with $c>1$ are
\be
\chi_{0}(\tau) = \frac{q^{-c/24}(1-q)}{\prod_{n=1}(1-q^{n})}\,,\qquad
\chi_{h}(\tau) = \frac{q^{h-c/24}}{\prod_{n=1}(1-q^{n})}\;\text{ for }
h>0\,,
\ee
where $q = e^{2\pi i \tau}$.
We can use the Dedekind eta function
\be
\eta(\tau) = q^{1/24}\prod_{n=1}(1-q^{n})
\,,\qquad
\eta(-\tau^{-1}) = (-i\tau)^{1/2} \eta(\tau)\
\ee
to define the \emph{reduced} characters
\be
\hat \chi_{h}(\tau) =: (-i\tau)^{1/4} \eta(\tau)\chi_{h}(\tau)
\en
and the reduced partition function
\be
\hat Z =: |\tau|^{1/2}|\eta(\tau)|^{2} Z(\tau,\bar \tau)
= \sum_{(h,\bar h)} N_{\bar h h}\,
 \overline{\hat \chi_{\bar h}(\tau)}\,\hat \chi_{h} (\tau)\, .
\ee
The function 
$|\tau|^{1/2}|\eta(\tau)|^{2}$ 
is invariant under $\tau\rightarrow -\tau^{-1}$,
so the reduced partition function
is invariant whenever the partition function is invariant.
The modular invariance condition~(\ref{eq:modinv}) becomes
\be
- P_{\odd} \hat Z_{0,0} = \sum_{(h,\bar h)\in S} N_{\bar h h} P_{\odd}\hat Z_{\bar h,h}
\,.
\label{eq:modinvred}
\ee
where
\be
\hat Z_{\bar h,h} = \overline{\hat \chi_{\bar h}(\tau)}\,\hat \chi_{h} (\tau)
= \left \{
\begin{array}{ll}
|\tau|^{1/2}\bar q^{-\gamma}(1-\bar q)q^{-\gamma}(1-q)\quad & 
\bar h=0,\; h =0 \\[1ex]
|\tau|^{1/2}\bar q^{\bar h-\gamma}q^{h-\gamma}(1-q) \quad & \bar h >0,\;h=0 \\[1ex]
|\tau|^{1/2}\bar q^{\bar h-\gamma}(1-\bar q)q^{h-\gamma} & \bar h =0,\;h>0\\[1ex]
|\tau|^{1/2}\bar q^{\bar h-\gamma}q^{h-\gamma} & \bar h>0,\; h >0
\end{array}
\right .
\ee
with
\be
\gamma = \frac{c-1}{24}
\,.
\ee
The advantage of this rewriting is
the relatively simple form of the functions $\hat 
Z_{\bar h,h}$.
The linear functional method applies as before if we redefine the 
cone $C_{S}$ as
\be
C_{S} = \bigg\{\sum_{(h,\bar h)\in S} N_{\bar h h} 
P_{\odd}\hat Z_{\bar h,h} \, : N_{\bar h h} \geq 0 \bigg\}
\ee
and the vector $v_{0}$ as
\be
v_{0} = - P_{\odd}\hat  Z_{0,0}\,.
\ee

\subsection{Differential operators}
Let us now discuss the space of linear functionals.
The partition function $\hat Z$ and all the functions $\hat Z_{\bar h,h}$ are real 
analytic functions of $\tau$,
so we can take our vector space of functions to consist of real analytic functions.
A complete set of linear functionals is given by evaluating
the Taylor series coefficients of a function at some fixed value of $\tau$.
The simplest choice is the self-dual value $\tau=i$, as adopted in \cite{Hellerman:2009bu}.
The linear functionals are then represented as real differential operators 
\be
\rho(f) = (\cD, f) =:  \cD  f\bigr|_{\tau=i}\,.
\ee
which are odd under $\tau\rightarrow -\tau^{-1}$.
We write the differential operators in the form
\be
\cD  = \sum_{j,k=0} d_{j,k} (\tau\partial_{\tau})^{j}
(\bar\tau\partial_{\bar\tau})^{k}
\,,\qquad d_{j,k}=\bar d_{k,j}
\label{eq:diffop}
\ee
so the oddness condition is simply
\be
d_{j,k}=0 \text{ for $j+k$ even}\,.
\ee
In practice, we solve the optimization problem~(\ref{eq:optprob1}), 
(\ref{eq:optprob2}) over 
finite dimensional subspaces of linear functionals given by the differential 
operators $\cD $ of order $\nD= 2n-1$.
As we increase the order $\nD=2n-1$,
the results become stronger, the 
bound more stringent.  The optimum result is 
obtained in the limit $\nD\rightarrow \infty$.

The advantage of this basis of linear functionals is that
the differential operators acting on the 
functions $\hat Z_{\bar h,h}$  become polynomials in $h$ and $\bar h$.
More precisely,
\be
(\cD, \hat Z_{\bar h,h}) = \left (\cD \hat Z_{\bar h,h}\right)\bigr|_{\tau=i}
= \hat p(h,\bar h)  e^{-2\pi(h+\bar h-2\gamma)} \,,\quad\text{for } h,\bar h > 0
\ee
where $\hat p(h,\bar h)$ is a polynomial in $h$ and $\bar h$ of the same 
degree as the differential operator.
The differential operators $\cD $ are in one-to-one correspondence with 
the polynomials $\hat p(h,\bar h)$.
Optimizing over 
differential operators $\cD $ is equivalent to optimizing over 
polynomials $\hat p(h,\bar h)$.
The oddness condition
on the differential operator translates into a collection of linear 
constraints on the coefficients of the polynomial.
Given the oddness condition on $\cD $, we have $\cD P_{\odd}\hat Z_{\bar h,h} = 
\cD \hat Z_{\bar h,h}$,
so the
semidefinite condition on the differential operator translates to
a semidefinite condition on the polynomial,
\be
\hat p( h,\bar h)\ge 0 \quad\text{ for $( h,\bar h) \in S$}
\label{eq:psemidef}
\ee
(with a small complication whenever a weight $(h,0)$ or 
$(0,\bar h)$ is in $S$ because of the extra factor of $1-q$ in $\hat Z_{\bar h,0}$
and $1-\bar q$  in $\hat Z_{0,h}$).

\subsection{Restriction to real $\beta$}
To make our computations more tractable, we introduce
two restrictions.
First, to avoid having to enforce the more complicated
semidefinite constraints $(\cD, \hat Z_{0,h})\ge 0$
and $(\cD, \hat Z_{\bar h,0})\ge 0$,
we specialize to CFTs that contain no representations with 
weights $(h,0)$ or $(0,\bar h)$,
as was done in \cite{Hellerman:2009bu}.
That is, we exclude CFTs that contain chiral fields besides 
the stress-energy tensor. 
A theory that did
contain such fields would have a symmetry algebra
bigger than just the Virasoro algebra. The proper way of dealing with 
such a theory
is to decompose the partition function into representations
of the larger symmetry algebra, which in general will lead to stronger 
constraints. 
Our discussion of the 
extended $N=(2,2)$ SCA in section~\ref{s:N2}
is an example of this procedure.

A more important issue is that we do not know an effective way to use the fact
that the spins $h-\bar h$ must be integers.
We do not know an effective way to enforce the semidefinite
condition $\hat p(h,\bar h)\ge 0$ 
on sets of the form $h+\bar h \ge \Delta_{1}$, $h-\bar h \in 
\Integers$.  
So, again following \cite{Hellerman:2009bu},
we limit ourselves to the 
subspace of linear 
functionals given by differential operators of the 
form
\be
\cD  = 
\sum_{k=0} d_{k} (\tau\partial_{\tau} + \bar 
\tau\partial_{\bar \tau})^{k}
\label{diffopimagtau}
\ee
subject to the oddness condition
\be
d_{k}=0 \text{ for $k$ even}\,.
\label{diffopimagtauoddness}
\ee
Equivalently,
we restrict the partition function to the imaginary
$\tau$ axis $\tau=i\beta$,
as a function of the real variable $\beta$,
\be
Z(\beta) = \tr \left ( e^{-\beta H}\right )\,,\qquad
H = 2\pi (L_{0}+\bar L_{0})\,.
\ee
The restricted partition function sees only the total weights 
$\Delta=h+\bar h$, and the differential operator~(\ref{diffopimagtau}) 
becomes 
\be
\cD  = \sum_{k=0} d_{k} (\beta 
\partial_{\beta})^{k}
\,.
\ee

Taking into account these two restrictions, 
the expansion of the reduced partition function in characters has the form
\be
\hat Z(\beta)
= \hat Z_{0}(\beta) + \sum_{\Delta\ge \Delta_{1}} N_{\Delta} \hat Z_{\Delta}(\beta)
\label{eq:realmodinv}
\ee
where
\be
N_{\Delta} = \sum_{h+\bar h=\Delta}N_{\bar h h}
\ee
is the multiplicity of irreducible representations with total 
weight $\Delta=h+\bar h$, and
\be
\hat Z_{\Delta}(\beta) = 
\left \{
\begin{array}{ll}
\beta^{1/2}q^{-2\gamma}(1-q)^{2},\quad & 
\Delta=0 \\[1ex]
\beta^{1/2}q^{\Delta-2\gamma},\;&\Delta >0
\end{array}
\right .
\label{eq:Zhatdelta}
\ee
with
\be
q = e^{-2\pi \beta}\,.
\ee
A differential operator $\cD$ corresponds to a polynomial
$\hat p(\Delta)$ by
\be
\label{eq:pofDelta}
(\cD, \beta^{1/2}q^{\Delta-2\gamma}) = 
\cD  \left (\beta^{1/2}q^{\Delta-2\gamma}
\right )\bigr|_{\beta=1} = \hat p(\Delta) e^{-2\pi(\Delta-2\gamma)}
\, .
\ee
It is crucial for our methods that
the polynomial now depends on one variable.
The semidefinite condition is
\be
\hat p(\Delta) \ge 0 \quad\text{for $\Delta \ge \Delta_{1}$}\, .
\ee
Note that the oddness and semidefinite conditions imply that the 
order $\nD$ of the differential operator  must be odd,
\be
\mathrm{ord}(\cD ) = \deg(\hat p)= \nD =  2n-1\,.
\ee
The objective that we want to to maximize is
\be
\cO = \rho(-v_{0}) = (\cD,  \hat Z_{0})
= \left[  \hat p(0)  - 2 \hat p(1)  e^{-2\pi}
+ \hat p(2)  e^{-4\pi} \right ] e^{4\pi\gamma}
\,.
\label{objphat}
\ee
The maximum, $\cO_{\max} $, is a monotonically increasing function of 
the gap $\Delta_{1}$,
because the semidefinite condition becomes weaker with increasing 
$\Delta_{1}$, so more differential operators are available in the 
optimization.
If $\cO_{\max} > 0$ then
no modular invariant partition function exists.
The upper bound $\bo$ on the gap is
given by the value of $\Delta_{1}$ where $\cO_{\max}= 0$.

\subsection{Explicit map to polynomials}
Having rewritten the optimization problem in terms
of polynomials, let us quickly give
explicit expressions for the conversion. This is a straightforward problem in linear
algebra.
For convenience we change to the variable $x = 2\pi(\Delta-2\gamma)$,
writing
\be
\label{eq:pofDeltatopofx}
\hat p(\Delta) =  p (x) = \sum_{k=0}^{2n-1} p_{k}x^{k}\,.
\ee
First, the semidefinite condition becomes
\be
p(x) \ge 0 \quad \forall 
x\ge x_{1}\,,\quad\text{where }
x_{1} = 2\pi (\Delta_{1}-2\gamma)\,.
\label{eq:positivity-p}
\ee
Next, the map from differential operators $\cD $ to polynomials $p(x)$ is
\be
\cD  \left (\beta^{1/2}e^{-\beta x}
\right )\bigr|_{\beta=1} = p(x) e^{-x}
\,.
\label{eq:pfromD}
\ee
We represent the differential operator and the polynomial as $2n$-vectors
\be
\vec d=(d_{0},d_{1},\,\ldots\,,d_{2n-1})\,,\qquad
\vec p=(p_{0},p_{1},\,\ldots\,,p_{2n-1})\,,
\ee
so the map from differential operator to polynomial is given by a matrix 
$G$,
\be
\vec p = G \, \vec d\,.
\label{eq:Gmatrix}
\ee
We compute $G$ in appendix~\ref{app:G}.
$G$ is upper triangular with diagonal entries $\pm 1$,
so is invertible.
This shows that optimizing over differential operators $\cD $ is indeed 
equivalent to optimizing over polynomials $p(x)$.
To express the oddness condition on $\cD $ as a condition on 
the polynomial $p(x)$, let $C$ be the 
$n \times 2n$ matrix that projects on the even coefficients 
of $\cD $,
\be
C\,\vec d =  (d_{0},d_{2},\,\ldots\,,d_{2n-2})  \,.
\ee
The oddness condition on $\cD $ becomes the $n$ linear 
conditions on $\vec p$
\be
C \, G^{-1} \, \vec p = 0\,.
\ee
Finally, the quantity (\ref{objphat}) we need to maximize is
(after dropping the positive factor $e^{4\pi \gamma}$)
\be
\cO = p(x_0)-2e^{-2\pi}p(x_0+2\pi)+e^{-4\pi}p(x_0+4\pi)
\label{eq:objp}
\ee
where
\be
x_{0}= -2\gamma\,.
\label{eq:x0}
\ee
The objective $\cO$ is a linear function of the coefficients of $p(x)$,
so we can write it as
\be
\cO = \vec o \cdot \vec p
\,
\ee
for some vector $\vec o$.

In summary, the optimization problem is to maximize the linear function
\be
\cO = \vec o \cdot \vec p
\label{eq:objpvec}
\ee
over polynomials $p(x)$ satisfying the semidefinite
condition~(\ref{eq:positivity-p}) and the $n$ linear conditions
expressing the oddness of the differential operator,
\be
C \, G^{-1} \, \vec p = 0\,.
\label{eq:constraintsp}
\ee

\subsection{Semidefinite programming (SDP)}\label{ss:SDPA}

We now need an effective way to scan over the space of
positive semidefinite polynomials.
For this we follow~\cite{Poland:2011ey}, expressing our optimization problem 
in the language of semidefinite programming (SDP).
The key step is to express the semidefinite condition~(\ref{eq:positivity-p}) on the polynomial 
$p(x)$ in terms of positive semidefinite matrices.
To this end we use the fact that any polynomial $p(x)$ which is nonnegative on the half-line $x\ge
x_{1}$ can be written in terms of sums of squares of polynomials 
\cite{Polya},
\be
p(x)  =  \sum_{a} q_{1,a}(x)^{2} + (x-x_{1}) \sum_{a} 
q_{2,a}(x)^{2}\,.
\ee
Equivalently, $p(x)$ can be written in terms of a pair of positive 
semidefinite matrices~\cite{Poland:2011ey}
\be
p(x)= \vec{x}^\transpose \, Y_{1}\, \vec{x}+ (x-x_1)\,\vec{x}^\transpose \, Y_{2} \, \vec{x}\,,
\label{eq:pY}
\ee
where
\be
\vec{x}= (1,x,x^2,\ldots,x^{n-1})
\label{eq:xvec}
\ee
and $Y_{1,2}$ are positive semidefinite $n{\times}n$ matrices.
Optimizing over polynomials $p(x)$ satisfying the semidefinite 
condition~(\ref{eq:positivity-p}) is equivalent to optimizing over 
the pair of positive semidefinite matrices $Y_{1,2}$.  
The objective function $\cO$ to be maximized
is a linear function~(\ref{eq:objpvec}) of $\vec p$ and therefore a 
a linear function of the matrix 
elements of  $Y_{1}$ and $Y_{2}$.
Likewise, the linear constraints~(\ref{eq:constraintsp}) become 
linear constraints on the $Y_{1,2}$.

Our optimization problem has now been expressed as a SDP problem:
maximizing a linear objective function over a set of semidefinite matrices
under a set of linear constraints.
Such problems have been well studied, and there
exist powerful SDP solvers.
We used the solver SDPA~\cite{SDPA}. For
details of our implementation of the SDP problem, see appendix~\ref{app:SDPA}.

\section{Virasoro symmetry}\label{s:Virasoro}
We first compute the bound $\Delta_{B}(c)$ as a function of $c$ for bosonic
conformal field theories with only Virasoro symmetry. 
The linear functional bound using differential operators 
$\cD $ of order $\nD= 3$ was analyzed in
\cite{Hellerman:2009bu} with the result
\be
\Deltamax = \frac{c}{6} + 0.47\ldots\,.
\ee
We want to see how much the bound can be lowered by going to higher order
differential operators.

By an argument of \cite{Friedan:2012jk},
there is no possibility of getting a linear functional bound smaller than $(c-1)/12$,
\be
\Deltamax \ge \frac{c-1}{12}\,,
\ee
because, for $\Delta_1<2\gamma$, there is no odd linear functional $\rho$ satisfying the 
semidefinite condition
\be
\rho(\beta^{1/2} e^{-2\pi \beta (\Delta-2\gamma)} ) \geq 0 \qquad \forall 
\Delta \geq \Delta_1\,,
\ee
\ie there is no hyperplane which has the cone $C_{\Delta_{1}}$ 
on one side of it.
So the linear functional method cannot exclude any $\Delta_{1}<  (c-1)/12$.
The proof that there is no such linear functional $\rho$ is given in appendix~\ref{app:bestbound}.

\subsection{Numerical results}

\begin{figure}[htbp]
\begin{center}
\includegraphics{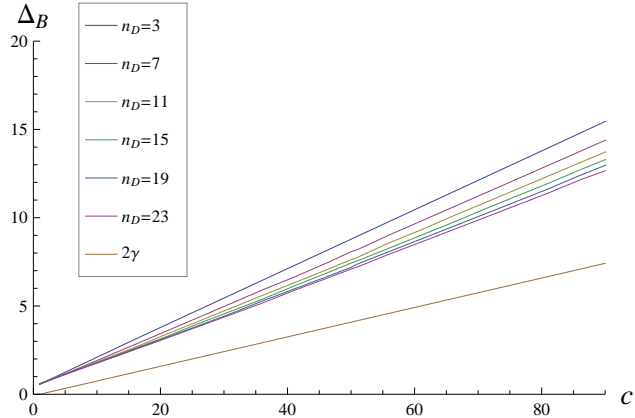}
\caption{$\Deltamax$ as a function of $c$. The bottom line is 
$2\gamma=(c-1)/12$,
which is
the smallest possible linear functional bound.}
\label{Fig:bosonLinear}
\end{center}
\end{figure}

In figure~\ref{Fig:bosonLinear} we plot the bound
$\Deltamax$ as a function of $c$ for various values of $\nD$,
the order of the differential operator.
The top line is the bound
obtained in \cite{Hellerman:2009bu}, with $\nD=3$.
The bottom line is $2\gamma=(c-1)/12$, which is
the smallest bound that the linear functional method could possibly 
produce.
We see that going to higher order differential operators
does improve the bound noticeably. It turns out that operators
of degree $\nD=4k+1$ never give significant improvements
over $\nD=4k-1$. This is because the optimal bound is given by 
$\Delta_{B}=\Delta_{1}$ when 
the vector $v_{0}$ lies exactly on the boundary of the 
cone $C_{\Delta_{1}}$.  The corresponding optimal polynomial $p(x)$ is of course
still nonnegative for $x\geq x_1$, but has double zeros.
For large $c$ $p(x)$ is essentially odd (see the next section
for a more precise statement), so the degree has to be $4k+1$.

\begin{table}
\begin{center}
\begin{tabular}{|c|c|c|c|}
\hline
$\nD$ & $c=1$ & $c=2$ & $c=50$ \\
\hline
3 & 0.615 & 0.788 & 8.8\\
7 & 0.604 & 0.748 & 8.07\\
11&0.604 & 0.741 & 7.63\\
15 &0.604 & 0.739 & 7.43\\
19& 0.603 & 0.739 & 7.19\\
23&0.603 & 0.739 & 7.09\\
27&0.603 & 0.739 &7.01\\
31&0.603 & 0.739 &6.92\\
35&0.603 & 0.739 &6.86\\
39&0.603 & 0.739 &6.81\\
43&0.603 & 0.739 &6.78\\
\hline
$2\gamma(c)$ & 0 & 0.0833 &  4.0833\\
\hline
\end{tabular}
\caption{Convergence of $\bo(c)$ as a function of the order $\nD=2n-1$ of the 
differential operator. The last row
is the smallest possible linear functional bound,
as discussed in section~\ref{s:Virasoro} and appendix~\ref{app:bestbound}.}
\end{center}
\label{t:conv}
\end{table}

To see the convergence of $\Deltamax$ as we increase $\nD$, we 
have tabulated 
$\Deltamax$ as a function of $\nD$ for $c=1,2$ and $50$ in 
table~\ref{t:conv}.\footnote{Strictly speaking our method does not apply for $c\leq 1$, since
then singular vectors appear in the Virasoro representations. When
we write $c=1$, what we mean is $c=1+\epsilon$ in the limit $\epsilon\rightarrow0$.}
For $c=1,2$ the bound converges very quickly, but stays far above
the theoretical minimum $2\gamma$.
For $c=50$ the bound converges much more slowly. Due to constraints on
our computation time we did not push beyond $\nD=43$. Our
data seems to show the $c=50$  bound converging geometrically 
to a value around $6.7$,
which is about halfway between the original bound of \cite{Hellerman:2009bu}
and the limiting value $2\gamma$.

\subsection{The large $c$ limit}

\begin{figure}[htbp]
\begin{center}
\includegraphics{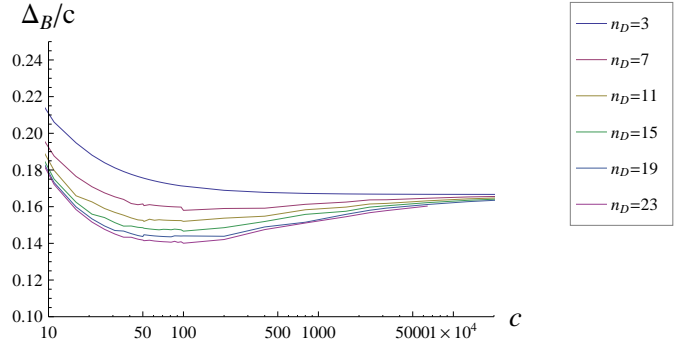}
\caption{The slope of $\Deltamax$ asymptotes to
$1/6$ for large $c$.}
\label{Fig:bosonLargec}
\end{center}
\end{figure}

Figure~\ref{Fig:bosonLinear} suggests that
$\Deltamax$ is almost a linear function in $c$.
We know from \cite{Hellerman:2009bu} that the $\nD=3$
bound goes as $c/6$ for large $c$,
and of course the lower limit  $2\gamma = (c-1)/12$ is linear in $c$. 
Since the new bounds are wedged between those
two, their leading behavior at large $c$ will also be linear.
The question is whether the slopes are smaller than
$1/6$.
In fact, the improved bounds all asymptote
to $c/6$, as can be seen in figure~\ref{Fig:bosonLargec}.

Let us try to understand this large $c$ behavior analytically. Since $c$ is the only parameter
in the problem, let us rescale $x=c y$, writing
$p(x) = q(y)$.
The objective function
(\ref{eq:objp})
is
\bea
\cO &=& q(y_0)-2e^{-2\pi}q(y_0+2\pi/c)+e^{-4\pi}q(y_0+4\pi/c)
\\
&=& (1-e^{-2\pi})^{2} q(y_0) + O(c^{-1}) \,,
\nonumber
\eea
where
\be
y_0=  - \frac1{12} + \frac1{12} c^{-1}\,.
\ee
To leading order in $c$ this 
means that $\cO$ vanishes when $q(y_0)$ vanishes.

Next note that according to appendix~\ref{sect:poddness} 
the oddness condition on the differential operator $\cD$ leads
to a relation between the even and 
odd parts of $q(y)$ under the reflection $y\rightarrow -y$,
\be\label{qe}
q_{\mathit{ev}} = \tanh(c^{-1}\Delta_{y})\,q_\mathit{odd}\,,\qquad 
\Delta_{y} = 
-\frac{1}{2}\frac{d}{dy}y\frac{d}{dy} 
=-\frac{1}{2}\left(y\frac{d^2}{dy^2}+\frac{d}{dy}\right)\,.
\ee
Thus, to leading order in $c$ the oddness condition is equivalent to
$q_{\mathit{ev}}(y)=0$.
So $q(y)$ is an odd function of $y$ up to $O(1/c)$ corrections.

A differential operator $\cD$ that solves the optimization problem 
will give $q(y)$ that is nonnegative for $y\ge y_{1}$ where 
$y_{1}=x_{1}/c$, 
$x_{1}= 2\pi(\Delta_{1}-2\gamma)$.
If this $\Delta_{1}$ is the bound $\Delta_{B}$,
then we also have that the objective $\cO$ is zero.

First, let us reproduce the $\nD=3$ large $c$ bound of \cite{Hellerman:2009bu}.
Parametrize the  third order differential operator
$\cD = (c^{-1}\beta\partial_\beta)^3 + A(c)c^{-1}\beta\partial_\beta$.
In the large $c$ limit, the map from differential operators to 
polynomials is
$(\beta\partial_\beta)^k \mapsto x^k$,
so
\be
q(y) = y^3+  A(c)y\,.
\ee
For the optimal differential operator --- the operator that gives the 
bound ---
the objective $\cO$ should vanish,
which in the large $c$ limit is the condition $q(y_{0})=0$.
So $A(c)$ must be $-y_0^2$ and
\be
q(y)= y(y-y_0)(y+y_0)\,.
\ee
This is nonnegative for $y\ge -y_0$, so we have $y_{1}=-y_{0}$,
which is
\be
x_1=2\pi(\bo-2\gamma)=-x_0 = 2\pi (2\gamma)
\ee
or
\be
\bo = 4\gamma = \frac{c}{6}+O(1)
\ee
which indeed reproduces the asymptotic result of \cite{Hellerman:2009bu}.

Let us now go to higher order differential operators.
Up to subleading contributions, we know that
$q(y)$ is odd in $y$,
and $q(y_0)=0$ for the optimal differential operator that gives the bound. 
It follows that $q(-y_0)=0$.
The asymptotic slope of the bound cannot be greater than 
$1/6$ because of the $\nD=3$ result.
So  $y_{1} \le -y_{0}$.
Therefore $q(y)$ is nonnegative for $y\ge -y_{0}$.
Now we only need to show that $y_{1}$ cannot be less than $-y_{0}$.
Then it will follow that $y_{1}=-y_{0}$ and 
we are done, getting again $x_{1} = -x_{0}$ which leads to
$\bo = \frac{c}{6}+O(1)$.

If the zero of $q(y)$ at $y=-y_{0}$ is of odd order, then, since 
$q(y)\ge 0$ 
for $y\ge -y_{0}$, we have $q(y)< 0$ for $y\lesssim -y_{0}$,
\ie for $y= -y_{0}-\epsilon$.
So  $y_{1}$ cannot be smaller than $-y_{0}$.  
The only way
out is if the zero has even order,
\be
q_{\mathit{odd}}(y) = y(y^2-y_0^2)^{2N}g(y)\,,
\ee
where $g$ is some even polynomial whose value at $-y_0$
is positive\footnote{To be slightly more precise,
we assume that the leading term of $g(y)$ in $c$ does not
vanish at $-y_0$. Otherwise we would absorb the root in the
prefactor.}.
Now we use (\ref{qe}) to calculate $q_{\mathit{ev}}(y)$ 
for $y\approx -y_0$
to $O(c^{-1})$,
\be
q_{\mathit{ev}}(y) = -\frac{1}{2c}(y^2-y_0^2)^{2N-2}
8N(2N-1)y_0^4\,g(y_0) \left[1+ O(y+y_0)\right]
\,.
\ee
This would make $q(y)$ go negative for $y$ slightly larger than $-y_{0}$,
which would imply $y_{1}>-y_{0}$, which we know is not true.
Therefore the zero must be of odd order and $y_{1}=-y_{0}$, giving 
$\bo = \frac{c}{6}+O(1)$.

Equation~(\ref{qe}) can be used to produce
a systematic expansion of $\bo(c)$ in powers of $1/c$.
We will not go into the details here, but only note that the first order 
correction is
\be
\bo(c)
= 
\frac{c}6 - \frac16 + \frac1{2\pi}
+ \frac{2}{e^{2\pi}-1} +O(c^{-1})
= \frac{c}6-0.00377+O(c^{-1})\,.
\ee

%%%%%%%%%%%%%%%%%%%%%%%%%%%%%%%%%%%%%%%%%%%%%%%%%%%%%%%%%%%%%%%%%

\section{$N=(2,2)$ superconformal theories}\label{s:N2}

Now we discuss $N=(2,2)$ superconformal theories,
which for example arise as nonlinear sigma models
on Calabi-Yau manifolds.
The central charge is $c=3d$ where $d$ is the complex 
dimension of the Calabi-Yau manifold.  We will study especially the 
case $d=3$.
The $N=2$ superconformal algebra first appeared in the context of 
string theory~\cite{Ademollo:1975an,Ademollo:1976pp}.
The unitary representations for $c\ge 3$ were classified in 
\cite{Boucher:1986bh}.
Their character formulas were derived in 
\cite{Dobrev:1986hq,Kiritsis:1986rv}.

\subsection{Generalized partition function}

Representations of the $N=2$ superconformal algebra 
are characterized by the eigenvalues of \emph{two} commuting 
operators:
the Virasoro generator $L_{0}$ and 
the generator $J_0$ of the $U(1)$ $R$-symmetry.
Abstractly, the Cartan 
algebra of the $N=2$ superconformal algebra is larger than that of 
the Virasoro algebra.
Each irreducible representation
is characterized by a weight $h$ and an integer charge $Q$.
The weight $h$ is the smallest eigenvalue of 
$L_{0}$.  The charge $Q$ is the eigenvalue of $J_{0}$ acting on the 
eigenspace $L_{0}=h$.
The characters
\eq
\tr \left (  q^{L_0-c/24} y^{J_0}\right )
\,,\qquad
q=e^{2\pi i \tau}\,,\quad y=e^{2\pi i z}\,,
\en
now depend
on an additional parameter $z$ conjugate to the 
conserved charge $J_{0}$.

The generalized partition function depends on parameters $z,\,\bar z$ 
in addition to the usual $\tau,\,\bar \tau$.
We will study the N-S partition function
\be
Z(\tau,\bar\tau,z,\bar z) = \tr \left (  q^{L_0-c/24}\bar q^{\bar L_0-c/24}
y^{J_0} \bar y^{ \bar J_0}\right )
\ee
where the trace is taken over the states of the NS sector.
We will use the invariance of the NS partition function under the S 
modular transformation \cite{Kawai:1993jk},
\be
S: (\tau, z) \mapsto (\tilde \tau,\tilde z)= (-1/\tau,z/\tau)\,,
\ee
\be \label{charTrafo}
Z(\tau,z) =  e^{-2\pi i \frac{d}{2} \frac{z^2}{\tau}} \,
e^{2\pi i \frac{d}{2} \frac{\bar z^2}{\bar \tau}}
\,Z(\tilde \tau,\tilde  z)\, .
\ee
We are now
writing $Z(\tau,z)$ instead of $Z(\tau,\bar\tau,z,\bar z)$
only to be succinct.

The NS sector is one of a continuum of sectors characterized by the 
monodromy of the charged fields around the spatial circle.
Spectral flow \cite{Schwimmer:1986mf} 
takes the NS sector to the 
other sectors of the theory, including the R 
sector,
so the modular transformation properties in the NS sector
imply the transformation properties in the other sectors.

Moreover, spectral flow  
implies that the theory contains an extended $N=2$ algebra,
which is generated by
the $N=2$ superconformal algebra plus two additional
holomorphic fields of conformal weight $d/2$.
These fields can be constructed from the $U(1)$ current $J$ as $e^{\pm \chi}$
where $J = \partial \chi$. They correspond to the spectral flow of 
the identity operator by $\pm 1$ periods.
The anti-holomorphic $N=2$ algebra is similarly extended.

It is advantageous to expand the partition function in characters of 
the largest algebra available.
The irreducible representations are bigger,
the Hilbert space decomposes into fewer irreducible 
representations, less multiplicity data is needed to specify the 
spectrum, and stronger constraints can be put on the spectrum.
So we expand in representations of the  extended $N=2$ algebra.

The representation theory
of the extended $N=2$ superconformal algebra
was analyzed in \cite{Odake:1988bh,Odake:1989dm,Odake:1989ev},
based on the representation theory of the unextended $N=2$ algebra.
Explicit formulas were derived for the characters of the irreducible 
representations of the extended algebra.
We will use only the NS representations and characters.
The character formulas are collected in appendix~\ref{app:SCA}.
We only quote the most important points here.

There are two kinds of irreducible representations:
the non-BPS or \emph{massive} representations,
and the BPS or \emph{massless} representations.
There are $d-1$ massive representations for each weight $h$,
subject to the unitarity constraint $h > \frac12 |Q|$.
The massive characters are
\be
\ma^Q_h(\tau,z):\qquad
\frac{3-d}{2} \le Q \le \frac{d-1}{2}\quad \text{for $d$ odd,}
\qquad
1 - \frac{d}{2} \le Q \le \frac{d}{2}-1\quad \text{for $d$ even.}
\ee
There are $d$ massless representations,
all having $h = \frac12|Q|$. 
The massless characters are
\be
\ml^Q(\tau,z):\qquad 
\frac{1-d}{2} \le Q \le \frac{d-1}{2}\quad \text{for $d$ odd,}
\qquad
1 - \frac{d}{2} \le Q \le \frac{d}{2}\quad \text{for $d$ even.}
\ee
We write $\mab_{h}$ for the $d-1$-vector with entries $\ma^Q_h$
and $\mlb$ for the $d$-vector with entries $\ml^Q$,
\be
\mab_{h} = (\ma^Q_h)\,,\qquad \mlb = (\ml^Q)\,.
\ee

The $N=(2,2)$ partition function decomposes into three parts,
\be \label{ZCY}
Z = Z_{\frac{1}{2}BPS} + Z_{\frac{1}{4}BPS} + Z_m
\ee
which come from tracing over three subspaces of the Hilbert space.
In terms of the characters,
\bea
Z_{\frac{1}{2}BPS} &=& \mlb^{\dagger}\, \Nb^{1/2}\, \mlb
\nonumber\\[1ex]
Z_{\frac{1}{4}BPS}&=&  \sum_{h} \mlb^{\dagger} \,\Nb_{h}^{1/4}\, \mab_{h}
+\sum_{\bar h} \mab_{\bar h}^{\dagger}\, \bar \Nb_{\bar h}^{1/4} \,\mlb 
\label{eq:Zincharacters}\\
Z_m &=& \sum_{h,\bar h }\mab_{\bar h}^{\dagger}\, \Nb^{m}_{\bar h h}\,  \mab_{h}
\,.
\nonumber
\eea
The four $\Nb$ matrices contain the multiplicities of the 
irreducible representations of the
extended $N=2$ algebras,
holomorphic and anti-holomorphic.

The $\frac{1}{2}$BPS part of the partition function comes from
the products of  left- and right-moving
BPS representations.
It is completely determined by the Hodge numbers $h^{ij}$ of the 
underlying Calabi-Yau manifold.
The $\frac{1}{4}$BPS part comes from 
the products of a BPS representation
with a massive representation, one left-moving, the other right-moving.
Part of the spectrum in these two subspaces is determined by the elliptic genus of the
Calabi-Yau manifold.
The third contribution
$Z_m$ comes from the subspace consisting of products of left- and right-moving massive
representations.  
In this subspace, the spectrum is a nontrivial quantum mechanical property of the
field theory,
determined by the geometry of the Calabi-Yau manifold. 

In our approach we assume that the topological part of the partition function is known,
the Hodge numbers and the elliptic genus.
We want to find constraints on the rest of the spectrum. 
Geometrically
this means that we start with a Calabi-Yau of fixed topology, and investigate
its (stringy) geometry.

Bounds on the gap $\Delta_{1}$ for $N=2$ theories were found in in \cite{Keller:2012mr} 
by considering low order differential operators in the two variables 
$\tau$ and $z$.
Here we simplify the problem considerably.
We express the partition function $Z(\tau,z)$ in terms of a 
matrix $\Mb(\tau)$
which is a real analytic function of $\tau$ alone,
and which transforms linearly under the $S$ modular transformation,
$\Mb(\tau) \rightarrow \Sb ^\dagger \Mb(\tilde \tau) \Sb $,
for a certain matrix of complex numbers $\Sb $.
Then we can obtain bounds
with the same techniques as in the bosonic case.

First we sketch the program for general $d$.
Then we specialize to $d=3$, which turns out to be
a considerably simpler special case.
We only carry out the program for $d=3$.

\subsection{Setup for general $d$}\label{sect:generald}

Following \cite{Odake:1989ev}, 
let us express the characters in a convenient basis.
Spectral flow implies that the characters
are quasiperiodic functions of $z$,
\be\label{quasiperiodic}
\ma^Q_h (\tau,z+\tau)=y^{-d}q^{-d/2}\ma^Q_h (\tau,z)\,, 
\qquad \ml^Q (\tau,z+\tau)=y^{-d}q^{-d/2}\ml^Q (\tau,z)\,.
\ee
Moreover the characters are periodic under $z\mapsto z+1$
and do not have any poles in $y$ away
from the origin and infinity.
Hermite's Lemma tells us that the space of such functions
has dimension $d$ over the functions of $\tau$.
One basis is given by the $d$ functions
\be
f_{d}^{Q}(\tau,z)= 
\frac{1}{\eta(\tau)}\sum_{m\in\Z}q^{\frac{d}{2}(m+Q/d)^2}y^{d(m+Q/d)}
\,,\qquad
f_{d}^{Q}=f_{d}^{Q+d}\,.
\ee
Any quasiperiodic function
can be written as a linear combination of the $f_{d}^{Q}$ with 
coefficients that are functions of $\tau$.
The $f_{d}^{Q}$ form a nice basis because they transform
linearly under the $S$ modular transformation.
Let $\fb_{d}$ be the $d$-vector with entries $f_{d}^{Q}$.
Then \cite{Odake:1989ev}
\be\label{ftrafo}
\fb_{d} (\tilde \tau,\tilde z) = e^{\frac{i\pi dz^2}{\tau}}\Sb _{d}\,\fb_{d} (\tau,z)
\,,\qquad
(\Sb _{d})_{Q'}^Q = d^{-\frac12} e^{-2\pi i QQ'/d}\, .
\ee
To get rid of the factor $e^{\frac{i\pi dz^2}{\tau}}$ in (\ref{ftrafo}),
we define
\be
\Fb_{d}= e^{\frac{i\pi dz^2}{2\tau}}\fb_{d}
\label{eq:Fddef}
\ee
which transforms by the numerical matrix $\Sb _{d}$,
\be
\label{Ftrafo}
\Fb_{d} (\tilde \tau,\tilde z) = \Sb _{d}\,\Fb_{d} (\tau,z)\,.
\ee
We use Hermite's lemma to expand the characters in the basis $f^{Q}$,
\eq
\label{eq:charactersfromfb}
\mab_{h} = \Gb^{\ma}_{h}(\tau) \fb_{d}
\,,\qquad
\mlb = \Gb^{\ml}(\tau) \fb_{d}
\en
where $\Gb^{\ma}_{h}$ is  a $d-1\times d$ matrix of functions of $\tau$,
depending on $h$,
and $\Gb^{\ml}$ is a $d\times d$ matrix of functions of $\tau$.

We use the above to rewrite the partition function.
First, in view of (\ref{charTrafo}), we define the reduced
partition function
\be
\hat Z(\tau,z) = \left| e^{\frac{i\pi dz^2}{2\tau}}
(-i\tau)^{1/4}\eta(\tau)\right|^2 Z(\tau,z)\,.
\label{eq:Neq2Zreduced}
\ee
$S$ modular invariance becomes simply
\eq
 \hat Z(\tau,z) = \hat Z(\tilde \tau,\tilde z) \,.
\label{eq:Neq2Zreducedmodinv}
\en
Using (\ref{eq:Zincharacters}), (\ref{eq:Fddef}), and~(\ref{eq:charactersfromfb}), 
we can write the reduced partition function as
\be\label{ZCYred}
\hat Z(\tau,z) = \Fb_{d}^\dagger \,\Mb (\tau)\,\Fb_{d}
\,,
\ee
where the $d\times d$ matrix $\Mb(\tau) $ is determined by the 
multiplicities,
\bea
\label{eq:MfromN}
\Mb &= &
(\hat \Gb^{\ml})^{\dagger} \, \Nb^{1/2}\, \hat \Gb^{\ml}
+ \sum_{h} (\hat \Gb^{\ml})^{\dagger} \,\Nb_{h}^{1/4}\,\hat \Gb^{\ma}_{h}
+ \sum_{\bar h} (\hat \Gb^{\ma}_{\bar h})^{\dagger} \, \bar \Nb_{\bar h}^{1/4} \,\hat \Gb^{\ml}
\\
\nonumber
&& \qquad {}+ \sum_{h,\bar h}(\hat \Gb^{\ma}_{\bar h})^{\dagger} \, \Nb^{m}_{\bar h h}\,\hat \Gb^{\ma}_{h}
\,.
\eea
with
\be
\hat \Gb^{\ml}(\tau) = (-i\tau)^{1/4} \eta(\tau) \Gb^{\ml}(\tau)
\,,\qquad
\hat \Gb^{\ma}_{h}(\tau) = (-i\tau)^{1/4} \eta(\tau) \Gb^{\ma}_{h}(\tau)
\,.
\ee
The crucial
point is that $\Mb(\tau)$ only depends on $\tau$, and no longer on
$z$.
All the dependence on $z$ is in the vector of functions $\Fb_{d}$.

Given the representation~(\ref{ZCYred}) of the reduced partition 
function and given the modular transformation properties~(\ref{Ftrafo})
of the vector of functions $\Fb_{d}$,
and the fact that the $F_{d}^{Q}$ are 
linearly independent as functions of $z$,
the modular invariance equation~(\ref{eq:Neq2Zreducedmodinv}) for the reduced partition function
is equivalent to the matrix equation
\be \label{matrixeq}
\Mb(\tau) = \Sb _{d}^\dagger\, \Mb(\tilde \tau)\, \Sb _{d}\,.
\ee
We are thus back at a variant of the bosonic modular invariance 
problem.

We apply the linear functional method.  
The function space is now the space of $d\times d$ matrices $\Ab(\tau)$ of functions of 
$\tau$, satisfying the oddness condition
\be
\Ab(\tau) = P_{\odd} \Ab(\tau) = \frac12
\left [\Ab(\tau) - \Sb _{d}^\dagger\, \Ab(\tilde \tau)\, \Sb _{d}\right]
\label{eq:Aoddness}
\ee
The linear functionals are represented by the $d\times d$ matrices 
$\cD^{QQ'}$ of differential operators in $\tau$ and $\bar \tau$
\eq
\cD = \Db(\tau\partial_{\tau})
\en
where $\Db$ is a matrix of polynomials in $\tau\partial_{\tau}$ and 
$\bar \tau\partial_{\bar \tau}$.
A matrix differential operator $\cD$ acts on a matrix of functions 
$\Ab$ by
\be
(\cD,\Ab) = \tr (\Db^\dagger\Ab)\bigr|_{\tau=i}\,.
\ee
Given the oddness condition~(\ref{eq:Aoddness}) characterizing the function space,
the linear functionals are given by
differential operators  satisfying the oddness condition
\be
\Db(\tau\partial_{\tau}) + \Sb \Db(-\tau\partial_{\tau}) \Sb ^\dagger  = 0\,,
\label{diffodd}
\ee
which is a straightforward matrix generalization of the oddness condition on the 
differential operators in the bosonic case,
and is easy to solve explicitly.
The $S$ modular invariance of $\Mb$ is now equivalent to
\be
\label{N2maineq}
(\cD,\Mb) = 0\quad \text{for all odd $\cD$}\,.
\ee
We separate $\Mb$ into two parts
\be
\Mb = \Mb_{0} + \Mb_{r}\,.
\ee
$\Mb_{0}$ comes from the multiplicities that we know, which includes
multiplicity 1 for the ground state representation plus the
multiplicities determined by the known topological properties of the 
Calabi-Yau manifold.  The rest of the multiplicities determine 
$\Mb_{r}$.  The semidefinite condition on $\cD$ is
\be
(\cD,\Mb_{r})\ge 0 
\label{eq:semidefinitegenerald}
\ee
for all possible multiplicities consistent with a given gap
$\Delta_{1}$.
Then, for all odd semidefinite $\cD$, modular invariance requires
\be
(\cD,\Mb_{0}) \le 0
\,.
\ee
We solve the optimization problem
\eq
\cO_{\max} = \max\{\,(\cD,\Mb_{0}): \cD \text{ odd 
semidefinite}\}\,.
\en
When the result is  $\cO_{\max}>0$, modular invariance is impossible 
with a gap equal to $\Delta_{1}$ or larger.

All that remains is to find effective ways to enforce the 
oddness and semidefinite conditions.
The matrix $\Mb_{r}(\tau)$ depends on the matrices of 
multiplicities by~(\ref{eq:MfromN}).  
It should be possible again to map the differential operators to 
polynomials.
But the map will involve
the differential operator $\cD$ acting on the 
change of basis matrices $\hat \Gb^{\ml}(\tau)$ and $\hat \Gb^{\ma}_{h}(\tau)$.
Explicit formulas can be derived for  $\hat \Gb^{\ml}(\tau)$ and 
$\hat \Gb^{\ma}_{h}(\tau)$, but the matrix entries will in general be infinite power series 
in $q$,
so the map to polynomials will not be simple.
The oddness condition on the polynomial will be ugly.

It turns out that the problem simplifies 
when $d=3$
because of a special property of the characters,
to the extent that we can do the numerical 
calculations using the same computer programs that we used for the bosonic 
case.

\subsection{$d=3$}\label{ss:d3}
Although we have set up our methods for the general case, we shall 
only apply them to Calabi-Yau threefolds. On the one hand, threefolds are of
most interest in string theory. On the other hand, several 
simplifications occur when $d=3$. 

We would like to make the entries of the  $\Mb_{r}(\tau)$ as simple
as possible. Ideally they should be simple monomials,
analogous to the Virasoro case. To this end let us
try to find a more appropriate basis.
For any $d$, the massive characters can be written \cite{Odake:1989ev}
\be\label{massf}
\ma^Q_h = \eta(\tau)^{-1} q^{h-\frac{d-1}{8}-\frac{Q^2}{2(d-1)}}
f_{1}^{0}f_{d-1}^{Q}\,.
\ee
The $d-1$ quasiperiodic functions  $f_{1}^{0}f_{d-1}^{Q}$
are transformed under $S$ 
by the matrix $\Sb_{d-1}$,
according to~(\ref{ftrafo}).
The problem is that
we need one more quasiperiodic function to form a basis
in which to expand the $d$ massless characters.
That last basis function will of course transform under S into
a linear combination of itself and the other $d-1$ basis functions,
but in general with coefficients that are 
power series in $q$. This means 
that the oddness condition on the matrix 
differential operators of section~\ref{sect:generald} will be
ugly (though most likely still possible to implement).

As it turns out, the situation 
for $d=3$ is much nicer.
The $d=3$ characters are
\be
\mlb = (\ml^{-1},\ml^{0},\ml^{1})\,,\qquad
\mab = (\ma^{0}_{h},\ma^{1}_{h})\,.
\ee
The massive characters are
\be
\ma^{0}_{h}= \eta^{-1}q^{h-\frac{1}{4}}
f_{1}^{0}f_{2}^{0}\,\qquad
\ma^{1}_{h} = \eta^{-1}q^{h-\frac{1}{2}}
f_{1}^{0}f_{2}^{1}\,.
\label{eq:massivechars}
\ee
The basis functions $f_{1}^{0}f_{2}^{Q}$ transform under S by
\be
\left (
\begin{array}{l}
f_{1}^{0}f_{2}^{0} \\
f_{1}^{0}f_{2}^{1}
\end{array}
\right ) (\tilde \tau,\tilde z) =e^{\frac{3\pi i z^{2}}{\tau}}
2^{-\frac12}\left(\begin{array}{cc} 1 &1\\
1&-1\end{array}\right)
\left (
\begin{array}{l}
f_{1}^{0}f_{2}^{0} \\
f_{1}^{0}f_{2}^{1}
\end{array}
\right ) (\tau,z) \,.
\ee
At the unitarity bound, $h=\frac12|Q|$, each massive representation  
decomposes into a sum of massless representations.
Thus two linear combinations of massless characters are given in 
terms of the basis functions $f_{1}^{0}f_{2}^{Q}$,
\be
\ma^0_0 = \ml^{-1} + \ml^0 + \ml^1
\,,\qquad
\ma^1_{1/2} = \ml^{-1} + \ml^1\,.
\ee
A third linear combination of the massless characters has a simple 
expression in terms of the $f_{3}^{Q}$
(derived in \cite{Odake:1989dm}, equations (2.13), (3.6), (3.17), and 
(3.18)),
\be
\ml^{1}-\ml^{-1} = f_{3}^{-}\,,\qquad f_{3}^{-}=f_{3}^{1} -f_{3}^{-1}\,.
\ee
The $S$ modular transformation takes the function $f_{3}^{1} 
-f_{3}^{-1}$ to itself,
\be
f_{3}^{-}(\tilde \tau,\tilde z) = e^{\frac{3\pi i z^{2}}{\tau}}(-i) 
f_{3}^{-}(\tau,z)
\,.
\ee
We take our basis functions to be $f_{1}^{0}f_{2}^{0}$, 
$f_{1}^{0}f_{2}^{1}$, and $f_{3}^{-}$.

Now we are in a position to 
simplify the $S$ modular invariance condition.
All of the characters are simple linear combinations of the three 
basis functions.
Substituting, the partition function becomes a sesquilinear 
expression in the basis functions.
The partition function now splits into four pieces:
(1) a piece proportional to $\bar f_{3}^{-}f_{3}^{-}$, 
(2) a piece proportional to just $f_{3}^{-}$,
(3) a piece proportional to just $\bar f_{3}^{-}$,
and (4) a piece containing neither $f_{3}^{-}$ nor $\bar f_{3}^{-}$.
Because the $S$ transformation does not mix $f_{3}^{-}$ with the other 
two basis functions,
each of these four pieces of the partition function must be 
separately invariant under S.

In fact, the middle two pieces of the partition function are identically zero.
Let us write the third piece $\bar f_{3}^{-} w(\tau,z)$.
Modular invariance of this piece requires that $w(\tau,z)$ transform 
in the same way as $f_{3}^{-}$.  
But is known that a function with such 
transformation properties
is unique: $f_{3}^{-}$ is the unique weak Jacobi form
of weight $0$ and index $3/2$ \cite{EichlerZagier,Gritsenko:1999fk}.
So $w(\tau,z)$ must be proportional to $f_{3}^{-}$.
To argue that that $w(\tau,z)=0$,
we use the expression for the $\frac12$BPS part of the partition 
function in terms of the Hodge numbers,
\bea
\label{eq:ZhalfBPS}
Z_{\frac12BPS}& =& \bar{\ml}^0\ml^0+ h^{1,1}( \bar{\ml}^{1}\ml^1 + \bar{\ml}^{-1}\ml^{-1})
+ h^{2,1} (  \bar{\ml}^{1}\ml^{-1}+\bar{\ml}^{-1}\ml^1) \\ 
& =& \ml^0\bar{\ml}^0+ \frac12 (h^{1,1}+h^{2,1} )|\ml^1 + \ml^{-1}|^{2}
+ \frac12 (h^{1,1}-h^{2,1} ) |\ml^1 - \ml^{-1}|^{2}\,.
\nonumber
\eea
$Z_{\frac12BPS}$ has no terms containing only one of $f_{3}^{-}$ and 
$\bar f_{3}^{-}$.
So $ w(\tau,z)$ must come entirely from the 
$\frac14$BPS representations, so the leading term in its 
$q$-expansion (the \emph{polar} part) vanishes, so $ w(\tau,z)$ must be 
identically zero.  By the same argument, the second piece of the 
partition function is zero.
So the partition function takes the form
\be
Z = Z'+ \frac12 (h^{1,1}-h^{2,1} ) |f_{3}^{-}|^{2}
\,.
\ee
where $Z'$ is sesquilinear in the two basis functions  $f_{1}^{0} f^{0}_{2}$ and $f_{1}^{0} 
f^{1}_{2}$.
The $|f_{3}^{-}|^{2}$ term is manifestly modular invariant,
so  $Z'$  must be 
modular invariant by itself.
To test modular invariance,
we can restrict our attention to $Z'$.

The above argument is exactly the argument of \cite{Keller:2012mr}
that the $\frac14$BPS representations do not contribute 
to the elliptic genus and are generically absent for $d=3$.
The elliptic genus is obtained by
flowing the partition function
to the R sector and then
taking the Witten index on the anti-holomorphic side.
The massive representations have zero index, so the elliptic genus 
comes entirely from the part of the partition function that 
contains $\bar f_{3}^{-}$, which carries index 2.  So the elliptic 
genus is $(h^{1,1}-h^{2,1} )f_{3}^{-} + 2 w(\tau,z)$.  By 
the argument given above, $w=0$.  So the elliptic genus comes entirely from the 
$\frac12$BPS representations.  Equivalently, the $\frac14$BPS representations all 
have index 0, so a generic perturbation of the field theory will lift 
them in pairs to massive representations.

It is an interesting question whether a similar simplification of the 
modular invariance condition can be found for $d\ne 3$.
We have some negative indications, although they are not definitive.
We checked for $d=2,4,5$ that there is no weak Jacobi form of weight 
0 and index $d/2$ that transforms into itself under $S$.
It might still be possible to find a function to complete the basis 
$f_{1}^{0} f^{Q}_{d-1}$ such that the $S$ transformation matrix is 
essentially numerical,
so the question remains open.

Now we specialize to the tractable case $d=3$.
We investigate
the modular invariance constraint on $Z'$, the part of the partition 
function sesquilinear in $f_{1}^{0} f^{0}_{2}$
and $f_{1}^{0} f^{1}_{2}$,
following the procedure outlined in section~\ref{sect:generald} above.
We change basis to
\be
\label{eq:Kdef}
\Kb =\left (
\begin{array}{l}
K^{0} \\
K^{1}
\end{array}
\right )\
= 
e^{\frac{i\pi dz^2}{2\tau}}
\left (
\begin{array}{l}
f_{1}^{0}f_{2}^{0} \\
f_{1}^{0}f_{2}^{1}
\end{array}
\right )
\ee
which simplifies the $S$ transformation to
\be
\label{eq:Smatrix}
\Kb(\tilde\tau,\tilde z) =\Sb \,\Kb(\tau,z)\,,\qquad
\Sb  =  \frac{1}{\sqrt{2}}\left(\begin{array}{cc} 1 &1\\
1&-1\end{array}\right)
\,.
\ee
The massive characters are  given in this basis by
\be
\label{eq:massivecharsinK}
\mab = \left (
\begin{array}{l}
\ma^{0} \\
\ma^{1} 
\end{array}
\right )
= 
e^{-\frac{i\pi dz^2}{2\tau}}
\eta(\tau)^{-1}
\left (
\begin{array}{cc}
q^{h-1/4} & 0 \\
0 & q^{h-1/2}
\end{array}
\right )
\Kb
\ee
and the massless characters by
\be
\label{eq:masslesscharsinK}
\mlb=
\left (
\begin{array}{l}
\ml^{-1} \\
\ml^{0} \\
\ml^{1}
\end{array}
\right )
=
e^{-\frac{i\pi dz^2}{2\tau}}\eta(\tau)^{-1}
\left (
\begin{array}{cc}
0 & \frac12 \\
 q^{-1/4} & - 1 \\
0 & \frac12
\end{array}
\right )
\Kb
+e^{-\frac{i\pi dz^2}{2\tau}}
\left (
\begin{array}{c}
-\frac12 \\
 0 \\
\frac12 
\end{array}
\right )f_{3}^{-}
\,.
\ee
We go to the reduced partition function, as in~(\ref{eq:Neq2Zreduced}),
\be
\hat Z' = \left| e^{\frac{i\pi dz^2}{2\tau}}
(-i\tau)^{1/4}\eta(\tau)\right|^2 Z'
\label{eq:Neq2Zprimereduced}
\ee
so the $S$ modular invariance condition becomes simply
\be
\hat Z'(\tau,z) = \hat Z'(\tilde \tau,\tilde z) \,.
\ee
$Z'$ has the form
\be
\tilde Z' = \Kb^{\dagger}\Mb(\tau) \Kb
\label{eq:Z'redinchars}
\ee
for $\Mb(\tau)$ a $2{\times}2$ matrix of functions of $\tau$ 
determined by the multiplicities.
We are re-using the notation $\Mb(\tau)$ despite the 
change in basis and the reduction in rank of the matrix from $d$ to 2.
The $S$ modular invariance condition is
\eq
\Mb(\tau) = \Sb ^\dagger \Mb(\tilde \tau) \Sb \,.
\label{eq:Mmodularinv}
\en
$\Mb(\tau)$ splits into a known part and an unknown part,
\be
\Mb(\tau) = \Mb_{0}(\tau) +\Mb_{r}(\tau) \,.
\ee
The known part comes from $Z_{\frac12BPS}$ as given by
equation~(\ref{eq:ZhalfBPS}).
It depends on the total Hodge number $ \htot$,
\be
\Mb_{0}(\tau) =  \Mb_{\frac12BPS}\\
=  |\tau|^{1/2}\left(\begin{array}{cc}\bar q^{-1/4} q^{-1/4}&-\bar q^{-1/4}\\
- q^{-1/4} &1+\frac{1}{2}{\htot}\end{array}\right)\,,
\qquad \htot = h^{1,1}+h^{2,1}\,.
\label{eq:M0}
\ee
The unknown part $\Mb_{r}(\tau)$ comes from  $Z_{\frac14BPS}+Z_{m}$.  
The $\frac14BPS$ sector makes no contribution to the elliptic genus,
so each $\frac14BPS$ representation in the Hilbert space contributes  one 
of four terms to the partition function: 
$\bar\ml^{0} \ma_{h}^{Q}$ or $\bar \ma_{\bar h}^{\bar Q} \ml^{0}$ 
or $(\bar\ml^{1}+\bar\ml^{-1}) \ma_{h}^{Q}$
or $\bar \ma_{\bar h}^{\bar Q}(\ml^{1}+\ml^{-1})$.
The latter two are simply massive representations at a unitarity 
bound,
\be
(\bar\ml^{1}+\bar\ml^{-1}) \ma_{h}^{Q} = 
\bar\ma_{1/2}^{1}\ma_{h}^{Q}\,,\qquad
\bar \ma_{\bar h}^{\bar Q}(\ml^{1}+\ml^{-1}) = \bar \ma_{\bar h}^{\bar 
Q}\ma_{1/2}^{1}\,.
\ee
The first two terms would cause trouble.
The identity $\ml^{0} = \ma^{0}_{0}-\ma^{1}_{1/2}$ means that a term 
$\bar\ml^{0} \ma_{h}^{Q}$ or $\bar \ma_{\bar h}^{\bar Q} \ml^{0}$ 
would make a negative contribution to $\Mb_{r}(\tau)$.
Our formulation of the semidefinite 
condition in the linear functional method requires that
all the  multiplicities appear in $\Mb_{r}(\tau)$ with the same 
sign.
So we make the assumption that there are no representations 
with characters $\bar\ml^{0}\ma^{Q}_{h}$ or $\bar \ma^{\bar Q}_{\bar 
h}\ml^{0}$.
This is exactly the assumption that the theory does not contain an
extended symmetry algebra -- i.e., no  holomorphic or anti-holomorphic 
fields besides the extended $N=2$ currents.
The ground state is in the representation with character $\ml^{0}$,
so such holomorphic or anti-holomorphic 
fields
would correspond exactly to representations with 
characters $\bar\ml^{0}\ma^{Q}_{h}$ or $\bar \ma^{\bar Q }_{\bar h}\ml^{0}$.

With this assumption, all the $\frac14BPS$ representations are just 
massive representations at a unitarity bound, so we can write
\bea
\Mb_{r}(\tau)&=&  |\tau|^{1/2} \sum_{h,\bar h}
\left(\begin{array}{cc}
 q^{\bar h-1/4} & 0 \\
0 & q^{\bar h-1/2}
\end{array}\right)^{\dagger}
\Nb^{m}_{\bar h h}
\;
\left(\begin{array}{cc}
 q^{h-1/4} & 0 \\
0 & q^{h-1/2}
\end{array}\right)
\\
&=& |\tau|^{1/2}\sum_{ h,\bar h}\left(\begin{array}{cc}
(N^{m}_{\bar h h})_{00}\,\bar q^{\bar h-1/4}q^{h-1/4}
&(N^{m}_{\bar h h})_{01}\,\bar q^{\bar h-1/4 } q^{h-1/2}\\[1ex]
(N^{m}_{\bar h h})_{10}\,\bar q^{\bar h-1/2}q^{h-1/4}
&(N^{m}_{\bar h h})_{11}\,\bar q^{\bar h-1/2}q^{h-1/2}
\end{array}\right)\,. \label{Mm}
\eea
where we extend the definition of the massive multiplicities 
$\Nb^{m}_{\bar h h}$ to
the unitarity bounds in order to include the paired $\frac14BPS$ representations,

\subsection{Linear functional method}

We continue to follow the procedure outlined in
section~\ref{sect:generald}, applying the linear functional method to
the modular invariance equation~(\ref{eq:Mmodularinv}).  For brevity,
we specialize from the beginning to $\tau=i\beta$, $\beta$ real,
though, as before, the general linear functional method takes the same
form for complex $\tau$.  The restriction to real $\beta$ is only
needed because of the limitations of our tools for expressing the
semidefinite condition on the linear functionals.

The linear functionals are the $2{\times}2$ matrices of differential 
operators
\be
\cD = \Db(\beta\partial_{\beta})
\ee
satisfying the oddness condition
\be
\Db(\beta\partial_{\beta}) + \Sb  
\Db(-\beta\partial_{\beta}) \Sb ^{\dagger} = 0
\,.
\ee
After a bit of algebra, the oddness condition can be expressed as
\bea
(\Db^{00}+\Db^{11})(\beta\partial_{\beta}) + (\Db^{00}+\Db^{11})(-\beta\partial_{\beta})
&=&0\,,\nonumber\\
(\Db^{00}- \Db^{11})(\beta\partial_{\beta}) + (\Db^{01}+\Db^{10})(-\beta\partial_{\beta})
&=&0\,,
\label{eq:N=2oddness}
\\
(\Db^{01}-\Db^{10})(\beta\partial_{\beta}) - (\Db^{01}-\Db^{10})(-\beta\partial_{\beta})
&=& 0\,.\nonumber
\eea
Given the restriction to real $\beta$, we could
set $\Db^{01}=\Db^{10}$ without loss of generality,
but it is not necessary to do so.

The semidefinite condition on $\cD$ is
\be
(\cD,\,\Mb_{r}) \ge 0
\ee
for all multiplicities $\Nb_{\bar h h}$ allowed by unitarity and by 
the gap condition $ h+\bar h\ge \Delta_{1}$.
Given the restriction to real $\beta$, we can
collapse the $2{\times}2$ matrix of multiplicities to functions of 
$\Delta=h+\bar h$,
\be
\Nb^{m}_{\Delta} = \sum_{h+\bar h=\Delta} \Nb^{m}_{\bar h h}\,.
\ee
The combined unitarity and gap conditions on the multiplicities are
\bea
(\Nb^{m}_{\Delta})_{00} &=& 0\,, \quad \Delta<\Delta_{1}\,,
\nonumber\\
(\Nb^{m}_{\Delta})_{01},(\Nb^{m}_{\Delta})_{10}  &=& 0\,, \quad 
\Delta<\max(\Delta_{1},1/2)\,,
\label{eq:Neq2multipliticityconstraints}\\
(\Nb^{m}_{\Delta})_{11} &=& 0\,, \quad \Delta<\max (\Delta_{1},1)
\,.\nonumber
\eea
The cone $C_{\Delta_{1}}$ in function space is the set of all 
matrices~(\ref{Mm}) where the $(\Nb^{m}_{\Delta})_{\bar Q Q}$ are 
allowed to range over all nonnegative \emph{real} numbers subject to 
the unitarity and gap conditions~(\ref{eq:Neq2multipliticityconstraints}).

We map each differential operator $\Db^{\bar QQ}$ to a polynomial
$p^{\bar QQ}(x)$
by equations~(\ref{eq:pofDelta}) and~(\ref{eq:pofDeltatopofx}).
The semidefinite condition on the $p^{\bar QQ}(x)$
can be read off from the monomials in (\ref{Mm})
and the conditions (\ref{eq:Neq2multipliticityconstraints})
on the multiplicities,
\bea
p^{00}(x) &\ge& 0\,, \quad x\ge2\pi(\Delta_{1}-1/2)\,,
\nonumber\\[1ex]
p^{01}(x),\,p^{10}(x) &\ge& 0\,, \quad 
x\ge2\pi\,\max(\Delta_{1}-3/4,\,-1/4)\,,\label{x1unitarity}
\label{eq:N=2semidefinite}
\\[1ex]
p^{11}(x) &\ge& 0\,, \quad x\geq2\pi\,\max(\Delta_{1}-1,\,0)
\nonumber
\,.
\eea
We can again use semidefinite programming, now with 4
semidefinite polynomials each
expressed in terms of a pair of positive semidefinite matrices.
The oddness condition~(\ref{eq:N=2oddness}) becomes a 
set of linear constraints on the vector of coefficients of the polynomials,
expressed in terms of the matrix $G$ of appendix~\ref{app:G}.

Finally let us turn to the objective function and the normalization.
For any semidefinite linear functional, modular invariance implies
\be
(\cD,\, \Mb_{0}) \le 0\,.
\label{eq:ineq}
\ee
The matrix $\Mb_{0}$ depends on the total Hodge number $\htot$.
For a fixed Hodge number we could proceed as in the bosonic problem,
maximizing the objective function $(\cD,\Mb_{0})$,
then solving for the value of $\Delta_{1}$ where the maximum crosses zero.
We would thus get an upper bound on the gap,
\be
\Delta_{1}\le \bo(\htot)
\,,
\ee
as a function of $\htot$.
Instead we follow a somewhat more efficient procedure.
We write
\be
\Mb_{0} = \Mb_{0'} + \htot \Mb_{h}\,,
\ee
where
\be
\Mb_{0'}=
|-i\tau|^{1/2}\left(\begin{array}{cc}\bar q^{-1/4} q^{-1/4}&-\bar q^{-1/4}\\
- q^{-1/4} &1\end{array}\right)
\,,\qquad
\Mb_{h}=
|-i\tau|^{1/2}\left(\begin{array}{cc}0&0\\
0 &\frac{1}{2}\end{array}\right)\,.
\ee
We choose our normalization condition to be 
\be
(\cD,\Mb_h) = 1
\label{eq:N=2normalization}
\ee
so that the inequality (\ref{eq:ineq}) becomes an upper bound on $\htot$,
\be
(\cD,\, \Mb_{0'}) + \htot \le 0\,.
\ee
For each value of $\Delta_{1}$,
we solve the optimization problem
\be
\cO = (\cD,\Mb_{0'})
\ee
\be
\cO_{\max} = 
\max_{\cD}\{\,\cO: \cD \text{ odd 
semidefinite}\}
\label{eq:N=2obj}
\ee
to get the lowest of these upper bounds on $\htot$,
\be
\htot \le  h^{\tot}_{B}(\Delta_{1})\,,\qquad h^{\tot}_{B} = - \cO_{\max}(\Delta_{1})\,.
\ee
This upper bound on $\htot$ must be a decreasing function of 
$\Delta_{1}$,
so is equivalent 
to an upper bound on $\Delta_{1}$ as a function of $\htot$.

\begin{figure}[htbp]
\begin{center}
\includegraphics{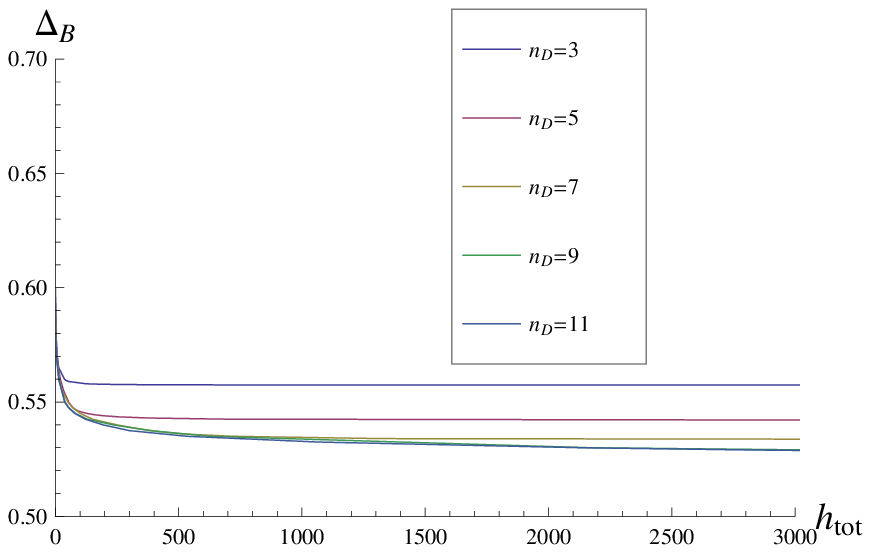}
\caption{$\bo(\htot)$ for various $\nD$.}
\label{Fig:N2diffops}
\end{center}
\end{figure}

In figure~\ref{Fig:N2diffops} we have plotted
the bound for various values of the order $\nD$
of the differential operator. 
As expected, 
$\bo$ is monotonically decreasing in $\htot$.
We find that $\bo$
converges very quickly in $\nD$ for small Hodge numbers.
The weakest bound is for vanishing Hodge numbers,
for which we find $\bo < 0.60$. Note in particular
that this means that the lowest lying state is always
a non-BPS state.

\subsection{Large Hodge numbers}\label{ss:direct}
We can see from figure~\ref{Fig:N2diffops}
that for fixed $n_D$ the bound becomes
slightly stronger with increasing total Hodge number.
Increasing $n_D$ improves the bound, the effect being
stronger for larger Hodge numbers.
Let us investigate the bound for very large Hodge numbers
a bit more carefully. 
The highest total Hodge number
for a Calabi-Yau known to exist at the moment 
is $\htot=491+11=502$ \cite{Kreuzer:2000xy,Taylor:2012dr}.
In fact, it is still an open question if the number of
topologically distinct CY threefolds is finite or not \cite{yau,Reid}.
For this reason it would be very interesting to find
a pathology in the spectrum for large enough Hodge numbers, such as
the bound $\bo$ becoming negative, hence ruling out
unitary sigma-models.
In fact, our methods cannot find any such pathology.
We show in this section that there is a linear functional bound 
$\bo=\frac12$ for asymptotically large $h_{\tot}$.
The argument is a variation on that of \cite{Keller:2012mr}
(correcting the result given there).
In the next section, we show that no linear functional bound can be 
lower than $\bo=\frac12$.
So the optimal linear functional bound in the limit 
$h_{\tot}\rightarrow\infty$ is $\bo=\frac12$,
which does not rule out any values of $\htot$.

To investigate large Hodge numbers it is useful to consider a 
particular set of linear functionals
\be\label{rhobeta}
(\rho_\beta,\Ab) := - P_{\odd}\Ab(\beta)_{00} =  - [ \Ab(\beta)  - \Sb^\dagger{\Ab(1/\beta)   }\Sb 
]_{00}\,.
\ee
Instead of Taylor expanding around $\beta=1$,
we evaluate the $(00)$ matrix element at some arbitrary $\beta$.
This clearly gives a linear functional $\rho_{\beta}$,
which moreover manifestly satisfies the oddness 
condition. To enforce the semidefinite condition we need to make sure that
$\rho_{\beta}$ is nonnegative on the cone $C_{\Delta_{1}}$.
Explicitly this means that $(\rho_\beta,\Mb_r) \ge 0$ for all matrices 
$\Mb_{r}$ of the form~(\ref{Mm}) when the $(\Nb_{\Delta})_{\bar Q Q}$ are 
allowed to range over all nonnegative \emph{real} numbers subject to 
the unitarity and gap conditions~(\ref{eq:Neq2multipliticityconstraints}).
Evaluating (\ref{rhobeta}) on such an $\Mb_{r}$, 
we have
\bea
\Mb_{00}(\beta) &=& \beta^{1/2} \sum_{\Delta}(N^{m}_{\Delta})_{00} 
\,e^{-2\pi\beta(\Delta -1/2)}
\\
{}[\Sb^\dagger{\Ab(1/\beta)}\Sb]_{00}
&=&
\frac12 \beta^{-1/2} \sum_{\Delta} \left[
(N^{m}_{\Delta})_{00} \, e^{-2\pi(\Delta-1/2)/\beta}
+ (N^{m}_{\Delta})_{01} \,e^{-2\pi(\Delta-3/4)/\beta}
\right .
\\
&&\qquad\qquad\quad
\left .
{}+ (N^{m}_{\Delta})_{10} \,e^{-2\pi(\Delta-3/4)/\beta}
+ (N^{m}_{\Delta})_{11} \,e^{-2\pi(\Delta-1)/\beta}
\right ]
\nonumber
\eea
so we can estimate
\be
(\rho_\beta,\Mb_r) 
\geq \sum_{\Delta} (N^{m}_{\Delta})_{00}\left(- \beta^{1/2} e^{-2\pi(\Delta-1/2)\beta}
+ \frac{1}{2}\beta^{-1/2} e^{-2\pi(\Delta-1/2)/\beta}\right)\,.
\ee
In particular this allows us to ignore any multiplicities other than 
$(N^{m}_{\Delta})_{00}$,
which is zero for $\Delta<\Delta_{1}$
and can take any nonnegative real value for $\Delta\ge\Delta_{1}$.
So $\rho_{\beta}$ satisfies the semidefinite condition if
\be
- \beta^{1/2} e^{-2\pi(\Delta-1/2)\beta}
+ \frac{1}{2}\beta^{-1/2} e^{-2\pi(\Delta-1/2)/\beta} \ge 0\,,\quad
\forall \Delta\ge\Delta_{1}\,.
\label{eq:betabound1}
\ee
Let us now see what values of $\beta$ we should choose.
For $\beta \ge 1$, the inequality is clearly not satisfied for large $\Delta$.
For $\beta<1$, (\ref{eq:betabound1}) is equivalent to
\be\label{betabound}
\Delta_{1} \geq \bo(\beta) =: \frac{1}{2} + \frac{\ln (2\beta)}{2\pi(\beta-\beta^{-1})}\,,
\ee
Note that $\bo(\beta)$ is monotonically decreasing in $\beta$, so we 
will want to take $\beta$ as large as possible
to get the lowest bound on $\Delta_{1}$.

The objective  we want to maximize is $(\rho_{\beta},\Mb_{0})$
where $\Mb_{0}$ is given by (\ref{eq:M0}),
\be
(\rho_{\beta},\Mb_{0})
= - \beta^{1/2} e^{\pi\beta}  +\frac{1}{2} \beta^{-1/2} 
\left [
(e^{\frac{\pi}{2\beta}}-1)^2 +\frac1{2}{\htot} 
\right ]
\,.
\ee
We estimate
\be
(\rho_{\beta},\Mb_{0})
> \frac{1}{4} \beta^{-1/2} ( \htot - 4\beta e^{\pi\beta})
\,.
\ee
If there is a value of $\beta$ such that $(\rho_{\beta},\Mb_{0})>0$,
then $S$ modular invariance is impossible
and the gap $\Delta_{1}$ can be excluded´.
So we can exclude $\Delta_{1}$ if
\be
\htot \ge 4\beta e^{\pi\beta}\,.
\ee
Define $\beta(h)$ as the solution to
\be
h = 4\beta e^{\pi\beta}\,.
\ee
In terms of the Lambert-$W$ function,
\be
\beta(h) = \frac{1}{\pi}W\left (\frac{\pi}{4}h\right )\,.
\ee 
We have $\rho_{\beta}$ semidefinite for $\Delta_{1}\ge \Delta(\beta)$ 
and we have $(\rho_{\beta},\Mb_{0})>0$ for $1<\beta< 
\beta(\htot)$, so we have a bound
\be
\bo(\htot) = \bo(\beta(\htot))
\ee
provided $\beta(\htot) >1$.  
It turns out that $\beta(\htot) > 1$ for $\htot\geq93$,
so this method does give a bound for
relatively large Hodge numbers.
On the other hand $\beta(\htot)$ grows monotonically
with $\htot$, so that in view of (\ref{betabound}) 
the bound becomes stronger and stronger
for larger Hodge numbers.
The Lambert $W$ function has an asymptotic expansion for large $z$ as
\be
W(z) = \ln z - \ln\ln z + o(1)\,,
\ee
so that for $\htot\rightarrow\infty$ we find $\bo=1/2$
(correcting the bound given in \cite{Keller:2012mr}).
Finally, as was shown in \cite{Keller:2012mr},
it follows from $(\rho_{\beta},\Mb_{0} +\Mb_{r})=0$ that
the number
of states below $\bo(\htot)$ grows linearly
in $\htot$ for large enough total Hodge number.

\begin{figure}[htbp]
\begin{center}
\includegraphics{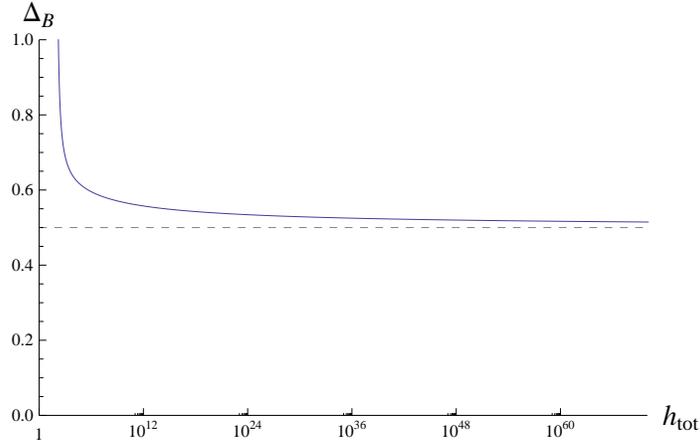}
\caption{The bound $\Delta_1(\htot)$ obtained from the Lambert $W$-function.}
\label{Fig:directMethod}
\end{center}
\end{figure}

\subsection{Best possible linear functional bound}
Finally let us show that our version of the linear functional method cannot
obtain a bound better than $\bo=\frac{1}{2}$.
This is essentially a repetition of the argument in the
Virasoro case. 
We show that, for $\Delta_{1}<\frac12$, 
there is no $2{\times}2$ matrix $\cD$ of differential operators
satisfying both the oddness condition and the semidefinite condition.
So the linear functional method with $\beta$ real cannot exclude any 
$\Delta_{1}<\frac12$.

Suppose there were such a matrix $\cD$.  Write $f_x = \beta^{1/2}e^{-\beta x}$.
Since $\Delta_{1}<\frac12$, the semidefinite condition~(\ref{eq:N=2semidefinite})
is
\bea
(\cD^{00},\,f_{x}) &\ge& 0\,,\qquad  x\ge 2\pi(\Delta_{1}-1/2)\,, \\
(\cD^{01},\,f_{x})\,, (\cD^{10},\,f_{x}) &\ge& 0\,,\qquad  x\ge 
2\pi(-1/4)\,, \\
(\cD^{11},\,f_{x}) &\ge& 0\,,\qquad  x\ge 0\,.
\eea
In particular, for all the matrix elements of $\cD$,
\be
(\cD^{{\bar Q Q}},\,f_{x}) \ge 0\,,\qquad  x\ge 0\,,
\label{eq:Dabnonneg}
\ee
and there exists $x'<0$ such that
\be
(\cD^{00},\,f_{x'}) \ge 0\,.
\label{eq:semi1}
\en
The gaussian integral identity~(\ref{eq:gaussianintegralidentity}) 
says that
\be
\tilde f_{x'} = \beta^{-1/2}e^{-\beta^{-1} x'}= \int_{0}^{\infty}dy\; 
B_{x'}(y) f_{y^2}\,\,,
\label{eq:gaussianintid}
\ee
for a certain function $B_{x'}(y)$ satisfying
\be
B_{x'}(y) >0 \qquad \textrm{for}\quad x'\le 0\,.
\ee
It then follows from~(\ref{eq:Dabnonneg}) that
\be
(\cD^{\bar Q Q},\,\tilde f_{x'})
= \int_{0}^{\infty}dy\; 
B_{x'}(y) (\cD^{\bar Q Q},\,f_{y^2})
\ge 0\,.
\label{eq:semi2}
\ee
The oddness condition says that, for any matrix $\Ab(\beta)$ of functions of 
$\beta$,
\be
(\cD,\Ab(\beta) + \Sb ^\dagger\Ab(\beta^{-1})\Sb )= 0\,.
\label{eq:bestboundoddness}
\ee
Let us take
\be
\Ab = \left(\begin{array}{cc} f_{x'} & 0 \\ 0 & 0 \end{array}\right)\,.
\ee
The oddness condition~(\ref{eq:bestboundoddness}) becomes
\be
(\cD^{00},\,f_{x'}) + 
\frac12(\cD^{00}+\cD^{01}+\cD^{10}+\cD^{11},\tilde f_{x'}) = 0\,.
\ee
Combined with the semidefinite conditions~(\ref{eq:semi1}) 
and~(\ref{eq:semi2}), this gives $(\cD^{\bar Q Q},\,f_{x'})=0$.
The gaussian integral identity~(\ref{eq:gaussianintid}) then says 
that $(\cD^{\bar Q Q},\,f_{x})=0$ for all $x\ge 0$,
and then that $(\cD^{\bar Q Q},\,f_{x})=0$ also for all $x<0$.
So $\cD=0$,
so there is no matrix $\cD$ of differential operators
satisfying both the oddness condition and the semidefinite condition,
so there is no possibility of a linear functional bound that excludes
any $\Delta_{1}<\frac12$.

Together with the results from section~\ref{ss:direct}, this shows that at least
for asymptotically large Hodge numbers we have a complete description
of the linear functional bound.

\medskip
\noindent {\bf Acknowledgment}

We thank S.~Hellerman, A.~Konechny, and C.~Schmidt-Colinet for useful
discussions, and S.~El-Showk for helpful discussions of the linear functional method.
This work was supported by the Rutgers
New High Energy Theory Center and by U.S. Department of Energy Grant No.~DE-FG02-12ER41813.

%%%%%%%%%%%%%%%%%%%%%%%%%%%%%%%%%%%%%%%%%%%%%%%%%%%%%%%%%%%%%%%%%%%%%%%%%%

\appendix
\appendixpage
\vskip2ex
\section{The matrix $G$}\label{app:G}
We calculate here the matrix $G$, defined by~(\ref{eq:pfromD}) 
and~(\ref{eq:Gmatrix}), which gives the map from the differential operator 
$\cD $ to the polynomial $p(x)$,
\be
\vec p= G\, \vec d\,,\qquad p_{j} = \sum_{k=0}^{\nD} G_{jk} d_{k}\,,
\ee
where
\be
p (x) = \sum_{k=0}^{\nD} p_{k}x^{k}
\,,\qquad
\cD  = \sum_{k=0}^{\nD} d_{k} (\beta 
\partial_{\beta})^{k}
\,,
\ee
and the map is
\be
\cD  \left (\beta^{1/2}e^{-\beta x}
\right )\bigr|_{\beta=1} = p(x) e^{-x}
\,.
\ee
The matrix elements $G_{jk}$ are given by the generating functional
\aeq{
\sum_{j,k=0} \frac1{k!} x^{j}G_{jk} t^{k} &=
\beta^{-1/2}e^{\beta x}
e^{t\beta\partial_\beta} \left (\beta^{1/2}e^{-\beta x}
\right )\bigr|_{\beta=1}
= e^{\frac12 t + x (1-e^{t})}
}
from which we see that
$G$ is upper triangular with diagonal matrix elements 
$G_{kk}=(-1)^{k}$
and is therefore invertible.

The inverse matrix, $G^{-1}$ reconstructs
the differential operator from the polynomial,
\be
\vec d = G^{-1}\,\vec p
\,,\qquad
d_{k} = \sum_{j=0}^{\nD} G^{-1}_{kj} p_{j}\,.
\ee
It can be calculated recursively.
The polynomials
\be
G^{-1}_j(t) = \sum_{k=0}^{\nD}  G_{jk} t^{k}
\ee
are defined by
\be
G^{-1}_j(\beta\partial_\beta) \left (\beta^{1/2}e^{-\beta x}
\right )\bigr|_{\beta=1} = x^{j} e^{-x} \,,
\ee
so $G^{-1}_0(t) = 1$ and, for $j>0$,
\aeq{
G^{-1}_{j}(\beta\partial_\beta) \left (\beta^{1/2}e^{-\beta x}
\right )\bigr|_{\beta=1}  &= x G^{-1}_{j-1}(\beta\partial_\beta) \left (\beta^{1/2}e^{-\beta x}
\right )\bigr|_{\beta=1}\\
&=\beta G^{-1}_{j-1}(\beta\partial_\beta) \beta^{-1}\left (-\beta\partial_\beta 
+\frac12\right )\left (\beta^{1/2}e^{-\beta x}
\right )\bigr|_{\beta=1}
}
so
\be
G^{-1}_{j}(t)  =  \left (-t +\frac12\right )G^{-1}_{j-1}(t-1)
= \prod_{m=0}^{j-1} \left (-t +\frac12-m\right )\,.
\ee

\section{Implementation of the SDP}\label{app:SDPA}
We use the SDP solver SDPA~\cite{SDPA}. 
It maximizes a linear function, the \emph{objective},
over a set of variables, which are positive semidefinite matrices,
subject to a set of linear constraints.

The variables are represented as a single positive semidefinite block diagonal matrix
$\Yb$, with prescribed block sizes.
In our case, for the Virasoro problem,
the variable matrix $\Yb$ consists of two $n{\times}n$ blocks
\be
\Yb = (Y_{1},Y_{2})\,,
\ee
corresponding to the matrices in (\ref{eq:pY}).
The linear functions are given by block matrices $\Fb$ with the same 
block structure as the variable matrix $\Yb$, acting by
\be
\Fb\bullet \Yb = \tr (\Fb^{\transpose} \Yb)
= \tr (F_{1}^{\transpose}Y_{1})+\tr (F_{2}^{\transpose} Y_{2})
\,.
\ee
The objective
\be
\cO = \Fb_{0}\bullet \Yb \,.
\ee
is calculated by
combining~(\ref{eq:objp}) and (\ref{eq:pY}).
The constraints
\be
\Fb_{i}\bullet \Yb = 0\,,\quad i=1,\ldots,n\,.
\ee
are calculated by 
combining~(\ref{eq:constraintsp}) and (\ref{eq:pY}).

Finally we need a normalization constraint because the optimization 
problem is invariant under a simple rescaling of the variable matrix 
$\Yb$(which is a rescaling of our differential operator $\cD$).
One possible normalization condition is
\be
\Fb_{n+1}\bullet \Yb = 1
\ee
where
\be
\Fb_{n+1} = (\mathbf{1}, \mathbf{1})
\,,\qquad
\Fb_{n+1}\bullet \Yb =\tr Y_{1}+\tr Y_{2}\,.
\ee
This is a robust normalization condition --- it singles out one point in 
each ray in the space of variables.

The block matrices $\Fb_{i}$ are the data that specify the SDP problem.
They depend on $\Delta_{1}$ via (\ref{eq:pY}).
For a given value of $\Delta_{1}$, we construct the SDP data using 
the symbolic mathematics program Sage \cite{SAGE}, which then hands off the SDP 
data to the extended-precision solver SDPA-GMP for solution.  
We found that extended precision arithmetic was needed to get reliable results.
The SDP solver returns an approximate maximum 
$\cO_{\max}$ and the corresponding solution matrix 
$\Ysol$.  All this is implemented as a Sage function 
$\Delta_{1}\mapsto (\cO_{\max},\Ysol)$.
We run the Sage 
root-finder on this function to find the solution of 
$\cO_{\max}(\Delta_{1}) =0$, to some 
specified accuracy.  The Sage root-finder 
reports a series of better and better approximate solutions.
Our bound $\Delta_{B}$ is the best approximate solution
for which $\cO_{\max}(\Delta_{B})>0$.

We can then verify the bound rigourously by working back 
from the SDPA solution matrix $\Ysol$ to calculate the 
odd differential operator $\Dsol $.  Then, from $\Dsol $, 
we calculate the polynomial $p(x)$ to verify the semidefinite 
condition (by finding all the roots of $p'(x)$),
and we calculate the objective $\rho(-v_{0})$ to verify 
its positivity.  All the verification calculations are done in extended precision 
arithmetic with enough precision to make rounding errors completely 
negligible.
When the verification succeeds, we have a specific differential 
operator $\Dsol$ satisfying the oddness and semidefinite conditions,
for which $\rho(-v_{0})>0$.
We therefore have a rigorous bound $\Delta_{B}$.

The implementation of the $N=2$ case is completely analogous.
We are optimizing over a $2{\times}2$ matrix of differential 
operators.  Each differential operator $\cD^{\bar Q Q}$ maps to a 
polynomial $p^{\bar Q Q}(x)$.  Each polynomial is 
represented by a pair of positive semidefinite matrices.
The variable matrix $\Yb$ now consists of 8 $n{\times}n$ blocks, 
$Y^{\bar Q Q}_{1,2}$.
The polynomials $p^{\bar QQ}(x)$ are given by
\be
p^{\bar QQ}(x) = \vec{x}^\transpose Y^{\bar Q Q}_{1} \vec{x} 
+ (x-x_1^{\bar QQ})\vec{x}^\transpose  Y^{\bar Q Q}_{2} \vec{x}\,,
\ee
where the $x_1^{\bar QQ}$ are obtained from (\ref{x1unitarity}).
The oddness condition (\ref{eq:N=2oddness}) 
leads to $4n$ constraint matrices $\Fb_i$.
The objective matrix $\Fb_{0}$ comes from~(\ref{eq:N=2obj}) and the 
normalization matrix $\Fb_{4n+1}$ from~(\ref{eq:N=2normalization}).
This normalization is strictly speaking not robust  --- there is a 
set of rays of measure zero in the variable space where the 
normalization condition has no solution.  In practice, this 
difficulty does not arise.
The $N=2$ program is simpler to execute.  For each value of 
$\Delta_{1}$, a Sage program provides the SDP data to the solver, 
which returns a bound on $\htot$.

\section{Modular transform of characters}

In this section, we
make two applications of the two-dimensional gaussian integral 
\be
\frac1\pi \int d^{2}y \;\beta^{1/2} e^{-\beta 
(y_{1}^2+y_{2}^{2})+ 2iy_1\sqrt{x}}
= \beta^{-1/2} e^{-\beta^{-1}x}
\,.
\label{eq:gaussianintegralidentity}
\ee

\subsection{Non-existence of semidefinite linear functionals
for $\Delta_{1} \le 2\gamma$}
\label{app:bestbound}

We rewrite (\ref{eq:gaussianintegralidentity}), setting $\vec y = 
(2\pi)^{1/2} \vec u$ and $x=2\pi(\Delta-2\gamma)$, to get
\be
2 \int d^{2}u \;\beta^{1/2} e^{-2\pi \beta 
(u_{1}^2+u_{2}^{2})+ 4\pi i u_1\sqrt{\Delta-2\gamma}}
= \beta^{-1/2} e^{-2\pi \beta^{-1}(\Delta-2\gamma)}
\label{eq:gaussian1}
\ee
which is a Fourier transform formula for the $S$ modular transformation
\be
\hat Z_{\Delta}(1/\beta) = 
2 \iint du_{1}du_{2} \; 
e^{4\pi i u_1\sqrt{\Delta-2\gamma}}
\;\hat Z_{2\gamma + u_{1}^2+u_{2}^{2}}(\beta)
\label{eq:fouriertransfreal}
\ee
of 
the reduced characters defined  (as functions of real $\beta$) by 
(\ref{eq:Zhatdelta}),
\be
\hat Z_{\Delta}(\beta) = \beta^{1/2} 
e^{-2\pi\beta(\Delta-2\gamma)}\,,\qquad \Delta>0\,.
\ee
Suppose  $\Delta_{1} \le 2\gamma$ and
suppose $\rho$ is a nonzero linear functional on the real analytic 
functions of $\beta$ satisfying the oddness condition and also the semidefinite condition 
\be
\rho(\hat Z_{\Delta}(\beta) ) \ge 0\,, \quad
\text{for } \Delta\ge\Delta_{1}\,.
\label{eq:rhosemidefinite}
\ee
From the oddness of $\rho$ and 
equation~(\ref{eq:fouriertransfreal})
we get
\be
\rho(\hat Z_{\Delta}(\beta)) 
= - \rho(\hat Z_{\Delta}(1/\beta)) 
= - 2 \iint du_{1}du_{2} \; 
e^{4\pi i u_1\sqrt{\Delta-2\gamma}}
\;\rho(\hat Z_{2\gamma + u_{1}^2+u_{2}^{2}}(\beta))
\,.
\label{eq:oddft}
\ee
The exponential in (\ref{eq:oddft}) is strictly positive
for any $\Delta$ in the range $\Delta_{1} \le \Delta \le 2\gamma$,
so (\ref{eq:oddft}) can only be consistent with~(\ref{eq:rhosemidefinite})
if $\rho(\hat Z_{\Delta}) =0$ for all $\Delta \ge 2\gamma$,
which by (\ref{eq:oddft}) implies  $\rho(\hat Z_{\Delta}) =0$ for all 
$\Delta$.
So $\rho=0$.
No such linear functional $\rho$ exists.

\subsection{Oddness condition on $p(x)$}
\label{sect:poddness}
The map~(\ref{eq:pfromD}) from differential operators $\cD $ to polynomials $p(x)$ is
\be
\cD  \left (\beta^{1/2}e^{-\beta x}
\right )\bigr|_{\beta=1} = p(x) e^{-x}
\,.
\ee
Under the modular transform $\beta\rightarrow\beta^{-1}$,
$p(x)$ goes to
\be
\tilde p(x) =  e^{x}\cD  \left (\beta^{-1/2}e^{-\beta^{-1} x}
\right )\bigr|_{\beta=1}
\,.
\ee
The oddness condition on $\cD$ is the condition on $p(x)$,
\be
p + \tilde p = 0\,.
\label{eq:poddness1}
\ee
Equation~(\ref{eq:gaussianintegralidentity}) gives the identity
\be
\tilde p(x) =  e^{x} \frac1\pi \int d^{2}y 
\;e^{-(y_{1}^2+y_{2}^{2})+ 2iy_1\sqrt{x}}\; p(y_{1}^2+y_{2}^{2}) 
\,.
\label{eq:pheatkern1}
\ee
We interpret the exponential in the integral in terms of the heat kernel for the two 
dimensional laplacian,
\be
\Delta_{2} = -\frac18 (\partial_{y_{1}}^{2}+\partial_{y_{2}}^{2})\,,
\ee
\be
e^{-2\Delta_{2}} f({\vec y}\,') = 
\frac1\pi \int d^{2}y
\;e^{-|\vec y - {\vec y}\,'|^{2}}
f(\vec y)\,.
\ee
Leting $f(\vec y)$ be the radially symmetric function
\be
f(\vec y) = p(|\vec y|^{2}) 
\,,
\ee
equation~(\ref{eq:pheatkern1}) becomes
\be
\tilde p(x) = e^{-2\Delta_{2}} f(i\sqrt{x},0)\,.
\label{eq:pheatkern2}
\ee
Given the radial symmetry of $f$, we can replace $\Delta_{2}$ with its 
radial part, and change variable from the radius $|\vec y|$ to $x=|\vec 
y|^{2} $, giving
\be
e^{-2\Delta_{2}} f(\vec y) = e^{-2\Delta} p(x)
\ee
where the radial part of $\Delta_{2}$ is the operator
\be
\Delta = - \frac12 \frac{d}{dx} x \frac{d}{dx} \,.
\ee
Equation~(\ref{eq:pheatkern2}) is now
\be
\tilde p(x) = e^{-2\Delta} p (-x)\,,
\ee
or
\be
\tilde p = R e^{-2\Delta} p
\ee
where $R$ is the reflection in $x$, acting on functions of $x$,
\be
Rp (x) = p(-x)\,.
\ee
The oddness condition on $p$, equation~(\ref{eq:poddness1}), is
$p + R e^{-2\Delta} p = 0$, or
\be
R p + e^{-2\Delta} p = 0\,.
\label{eq:oddnessDelta1}
\ee
If we separate $p(x)$ into its even and odd parts under the reflection $x\rightarrow -x$,
\be
p_{\mathit{ev}} =\frac12 (1+R) p\,,\qquad
p_{\mathit{odd}}=\frac12 (1-R) p\,,
\ee
the oddness condition (\ref{eq:oddnessDelta1}) on $p$ becomes
\be
p_{\mathit{ev}} - p_{\mathit{odd}} + e^{-2\Delta} (p_{\mathit{ev}} + p_{\mathit{odd}}) = 0\
\ee
or
\eq
p_{\mathit{ev}} = \tanh(\Delta)\, p_{\mathit{odd}}\,.
\en

\section{Representations of the extended $N=2$ SCA}\label{app:SCA}\label{app:N2reps}
A worldsheet theory for Calabi-Yau compactification of
type II superstring theory without flux or for $N=(2,2)$ compactification
of heterotic string theory 
has $N=2$ superconformal symmetry.
In addition the theory is invariant under spectral flow.
Spectral flow by one unit maps
the Neveu-Schwarz (NS) sector to itself,
taking the $N=2$ symmetry algebra to a set of additional operators 
that extend the symmetry algebra.
This is equivalent to the geometric fact that a CY $d$-fold always 
carries
a holomorphic $(d,0)$ form.
The Hilbert space decomposes into irreducible representations of the
extended $N=2$ superconformal algebra.
The representation theory
of the extended $N=(2,2)$ superconformal algebra
has been studied in \cite{Odake:1988bh,Odake:1989dm,Odake:1989ev},
whose main results we repeat here,
especially
\cite{Odake:1989ev}, equations (3.2), (3.3), (4.8), (4.9), 
and (4.11).

For a Calabi-Yau $d$-fold the extended $N=2$ superconformal
algebra has central charge $c=3d$. Let us define
$k=d-1$.
The irreducible representations are characterized by the eigenvalues
$(h,Q)$ of the generators $L_{0}$ and $J_{0}$ acting on the lowest 
weight subspace of the representation.
The character of a representation is
\eq
\tr \left (  q^{L_0-c/24} y^{J_0}\right )\,,\qquad
q=e^{2\pi i \tau}\,,\quad y=e^{2\pi i z}\,.
\en
Define
\be
F_{NS}(\tau,z)=\prod_{n\geq1}\frac{(1+yq^{n-1/2})(1+y^{-1}q^{n-1/2})}{(1-q^n)^2}
= \frac{q^{\frac18}}{\eta(\tau)^{3}} \sum_{m\in\Integers} q^{\frac12 m^{2}}y^{m}
\ee
which is the character of the unextended $N=2$ algebra
without any relations (\ie the character of the $h=0$, $Q=0$ Verma module).

Define the functions
\be
f_{d}^{Q}(\tau,z)= 
\frac{1}{\eta(\tau)}\sum_{m\in\Z}q^{\frac{d}{2}(m+Q/d)^2}y^{d(m+Q/d)}
\,,\qquad
f_{d}^{Q}=f_{d}^{Q+d}\,.
\ee
In this notation,
\be
F_{NS}(\tau,z)= q^{\frac18}\eta(\tau)^{-2} f_{1}^{0}(\tau,z)\,.
\en

\vskip1ex
\noindent
{\bf Massive representations}

There are $d-1$ massive representations for each $h$, subject to the 
unitarity condition $h > \frac12 |Q|$.  $Q$ can take one of $d-1$ 
integer values:
\be
\frac{3-d}{2} \le Q \le \frac{d-1}{2}\quad \text{for $d$ odd,}
\qquad
1 - \frac{d}{2} \le Q \le \frac{d}{2}-1\quad \text{for $d$ even.}
\ee
The massive characters are
\bea
\ma^Q_h(\tau,z) &=& q^{h -c/24}
F_{NS}(\tau,z)q^{-Q^2/2k}\sum_{m\in\Z}q^{\frac{k}{2}(m+Q/k)^2}y^{k(m+Q/k)}\,.
\\
&=& \eta(\tau)^{-1} q^{h -k/8-Q^2/2k} f_{1}^{0}(\tau,z) 
f_{k}^{Q}(\tau,z)\,.
\eea
Note that 
\be
\ma^{-Q}_h(\tau,z) = \ma^Q_h(\tau,-z)\,.
\ee

\vskip1ex
\noindent
{\bf Massless representations}

There are $d$ massless representations, all at the unitarity bound 
$h=\frac12|Q|$.  $Q$ can take one of $d$ 
integer values:
\be
\frac{1-d}{2} \le Q \le \frac{d-1}{2}\quad \text{for $d$ odd,}
\qquad
1 - \frac{d}{2} \le Q \le \frac{d}{2}\quad \text{for $d$ even.}
\ee
The massless characters for $Q>0$ are
\be
\ml^Q(\tau,z) = q^{\frac12 Q -c/24} F_{NS}(\tau,z)q^{-Q^2/2k}\sum_{m\in\Z}
\frac{q^{\frac{k}{2}(m+Q/k)^2}y^{k(m+Q/k)} }{1+yq^{m+1/2}}\,.
\label{eq:masslesscharsQpos}
\ee
For $Q<0$, the massless characters are given by
\be
\ml^Q(\tau,z) = \ml^{-Q}(\tau,-z)\,.
\ee
For $Q=0$, the massless character is
\be
\ml^0(\tau,z) = q^{-c/24}F_{NS}(\tau,z)\sum_{m\in\Z}
\frac{(1-q)q^{\frac{k}{2}m^2}y^{km}}{(1+yq^{m+1/2})(1+y^{-1}q^{-m+1/2})}\,.
\ee
Note that, in equation~(\ref{eq:masslesscharsQpos}), substituting 
$Q\rightarrow d-Q$, $z\rightarrow -z$, $m\rightarrow -m-1$ gives the 
identity
\be
\ml^{d-Q}(\tau,z)=\ml^{Q}(\tau,-z)\,,
\ee
so, for $Q<0$, the massless character is $\chi^{d+Q}$,
so we could label the massless representations by $Q=0,1,\ldots,d-1$
with characters $\ml^{Q}$ given by 
equation~(\ref{eq:masslesscharsQpos}).
Note, however, that $Q$ loses its meaning if we do this.

When the massive representations reach the unitarity
bound $h=\frac{1}{2}|Q|$, they become reducible and decompose
into massless representations as
\be
\ma^Q_{|Q|/2} =
\left\{\begin{array}{ll}
\ml^Q+\ml^{Q+1} & Q > 0\,,\\
 \ml^0 +\ml^1+\ml^{-1}&Q = 0 \,,\\
\ml^{Q}+\ml^{Q-1} & Q<0\,.
 \end{array}\right.
\ee

The Witten index of the massive representations is zero.
The index of the massless representations is (after spectral flow to the 
R sector)
\be
\mathit{ind}(\ml^{Q}) = \left\{\begin{array}{ll}
(-1)^{d-Q}& Q > 0\,,\\
1+(-1)^d&Q = 0\,, \\
(-1)^{-Q}& Q<0\,.
 \end{array}\right.
\ee

%%%%%%%%%%%%%%%%%%%%%%%%%%%%%%%%%%%%%%%%%%%%%

\end{document}